\documentclass[traditabstract]{aa}  

\pdfoutput=1
\usepackage{graphicx}
\usepackage{txfonts}
\usepackage[authoryear]{natbib}
\usepackage{longtable}
\usepackage{supertabular}
\usepackage{pdflscape}

\begin{document}

\title{The Hubble Legacy Archive ACS Grism Data}

\titlerunning{HLA ACS Grism Data}

\author{
M.\ K\"ummel
\and
P.\ Rosati
\and
R.\ Fosbury
\and
J.\ Haase
\and
R.N.\ Hook
\and
H.\ Kuntschner
\and
M.\ Lombardi
\and
A.\ Micol
\and
K.K.\ Nilsson
\and
\\
F.\ Stoehr
\and
J.R.\ Walsh
}
\authorrunning{K\"ummel et al.}

\institute{Space Telescope -- European  Coordinating Facility,
Karl-Schwarzschild-Str. 2, D-85748 Garching, Germany \email{mkuemmel@eso.org}}
\date{v1.2 - To be submitted Dec 30, 2010; accepted ...}

\abstract{
A public release of slitless spectra, obtained with ACS/WFC and the G800L 
grism, is presented. Spectra were automatically extracted in a uniform way 
from 153 archival fields (or "associations") distributed across the two
Galactic caps, covering all observations to 2008. 
The ACS G800L grism provides a wavelength range of 
$0.55$--$1.00 \mu$m, with a dispersion of $40 \ \AA / \mbox{pixel}$ 
and a resolution of $\sim 80\ \AA$ for point-like sources. The ACS G800L 
images and matched direct images were reduced with an automatic pipeline
that handles all steps from archive retrieval, alignment and astrometric
calibration, direct image combination, catalogue generation, spectral 
extraction and collection of metadata. The large number of extracted
spectra ($73,581$) demanded automatic methods for quality control 
and an automated classification algorithm was trained on the visual 
inspection of several thousand spectra. The final sample of quality 
controlled spectra includes $47,919$ datasets (65\% of the total number 
of extracted spectra) for $32,149$ unique objects, with a median 
$i_{\rm AB}$-band magnitude of 23.7, reaching 26.5 AB for the faintest 
objects. Each released dataset contains science-ready 1D and 2D 
spectra, as well as multi-band image cutouts of corresponding sources
and a useful preview page summarising the direct and slitless data,
astrometric and photometric parameters. This release is part of the 
continuing effort to enhance the content of the Hubble Legacy Archive 
(HLA) with highly processed data products which significantly facilitate 
the scientific exploitation of the Hubble data. In order to characterize
the slitless spectra, emission-line flux and
equivalent width sensitivity of the ACS data were compared with public 
ground-based spectra in the GOODS-South field. An example list of emission line 
galaxies with two or more identified lines is also included, covering 
the redshift range $0.2-4.6$. Almost all redshift determinations outside
of the GOODS fields are new. The scope of science projects possible with 
the ACS slitless release data is large, from studies of Galactic stars 
to searches for high redshift galaxies.
} 

\maketitle

\section{Introduction}
\label{intro}
Slitless spectroscopy is primarily a survey tool and most of the use
on HST has been with this aim. On account of the lack of selection of
target(s) by a slit, then the spectra of all objects in a given field
are recorded, implying that the range of targets can be large - from
the nearest stars to the most distant galaxies. The primary
disadvantages of slitless spectroscopy in comparison with slit
spectroscopy are the mutual contamination of spectra, the higher sky
background (as compared to imaging) and
the range in spectral resolution dictated by the size of the
dispersing objects ('virtual slits'). Since early in its history HST
has been equipped with instruments capable of slitless spectroscopy
\citep[for an overview see][]{Walsh2010}, but the most-used
modes have been on NICMOS and ACS and, since Servicing Mission 4,
WFC3.

As part of the Hubble Legacy Archive (HLA) project to create
well-calibrated science data and make them accessible via
user-friendly archives, the Space Telescope European Coordinating Facility
(ST-ECF) has exploited its expertise in
slitless spectroscopy to uniformly extract spectra from widely used
slitless spectroscopy modes and to serve the high level science data products
through an archive. The HLA is a collaboration between Space Telescope
Science Institute (STScI), the Canadian Astronomy Data Centre (CADC)
and the ST-ECF. In
the first part of this project about $2\,500$ spectra from the NICMOS G141
grism, covering the wavelength range 1.10 to 1.95$\mu$m, were
extracted and made publicly available
({\tt http://hla.stsci.edu/STECF.org/archive/hla/}).
%http://hla.stecf.org/
The process of extraction, the limitations
of the data served and examples of a few of the spectra are described
in \cite{Freudling08}.  In this paper we present the
second part of that project in which all the ACS G800L slitless data
taken with the Wide Field Channel (WFC) were extracted and the data
products released through the HLA. The ACS G800L grism can also be
used with the High Resolution Channel (HRC) and a separate release of
slitless spectroscopy in the Hubble Ultra Deep Field was made
%\citep[see][and {\tt http://www.stecf.org/UDF/HRCpreview.html}]{Walsh2004},
\citep[see][and {\tt http://archive.stsci.edu/prepds/udf/stecf\_udf/} {\tt HRCpreview.html}]{Walsh2004},
but is not considered here.

The bulk (76\% by number of pointings, 70\% by exposure time) of the ACS 
G800L spectroscopy was taken in four programmes,
two of which were pure parallel. Table \ref{tab:progover} presents
an overview by proposal of the contents of the HLA ACS slitless 
spectra, listing the number of associations and grism
exposure time. The ACS Pure Parallel Ly$\alpha$
Emission Survey (APPLES, Proposal No. 9482, PI: J.\ Rhoads)
used the $\geq$3 orbit parallel opportunities to obtain a
direct image and multiple grism images. Depending on the length of the
parallel opportunity, \texttt{F850LP}, \texttt{F606W} and \texttt{F435W}
images were added to the core \texttt{F775W} direct image.
Low Galactic latitude pointings
($|b|>20^\circ$) were avoided. A very similar parallel
programme -- the ACS Grism Parallel Survey of Emission-line Galaxies at
Redshift z$<$7 -- was also performed in HST Cycle 11
(9468, PI: L.\ Yan). 
Results from some deeper pointings ($>$ 12ks) were reported
for 11 high Galactic latitude fields; 601 compact emission-line
galaxies at $z \leq 1.6$ were found from H$\alpha$, H$\beta+$[O~III]
and [O~II] lines \citep{Drozdovsky}.

The Grism-ACS Program for Extragalactic Science (GRAPES, 9793,
PI: S.\ Malhotra) was a primary programme to obtain deep grism exposures in
the Hubble Ultra Deep Field (HUDF). The strategy and data reduction is fully
described in  \cite{Pirzkal2004} and a deep G800L image from another primary
programme in the same
field was included in the analysis
(proposal 9352, PI: M.\ Riess; see Riess et al., 2004 for
results on a supernova at redshift $z=1.3$). The strategy adopted for
GRAPES was to use the very deep HUDF imaging to provide the positions of 
the sources to be extracted, and to use comparably
short (700s) direct image exposures in \texttt{F606W} to align the grism images
to the HUDF catalogue. A range of science was published from these data
including detections of faint Galactic stars \citep{Pirzkal2005} and a
grouping of high-z galaxies \citep{Malhotra2005}.  A set of 1400
extracted spectra (with $i < 27.0 \mbox{ mag}$) are publicly available at
{\tt http://archive.stsci.edu/prepds/grapes/}

The GRAPES observations were extended to larger fields with the PEARS
programme (Probing Evolution And Reionization Spectroscopically,
proposal 10530, PI: S.\ Malhotra), including the Hubble Deep Field
North (HDFN) as well as the Chandra Deep Field South (CDFS). Short
\texttt{F606W} images were employed to match the coordinates to the Great
Observatories Origins Deep Survey (GOODS) catalogue \cite{Giava04}.  
Results on emission line galaxies \cite{Straughn}, early type 
galaxies \cite{Ferreras2009} and faint stars \cite{Pirzkal2009} have been 
published. Spectra from PEARS
($4,082$ for the CDF-N field and $5,469$ for the CDF-S) are
publicly available through a searchable interface at
{\tt http://archive.stsci.edu/prepds/pears/}. 

\begin{figure}[t]
\includegraphics[width=\hsize]{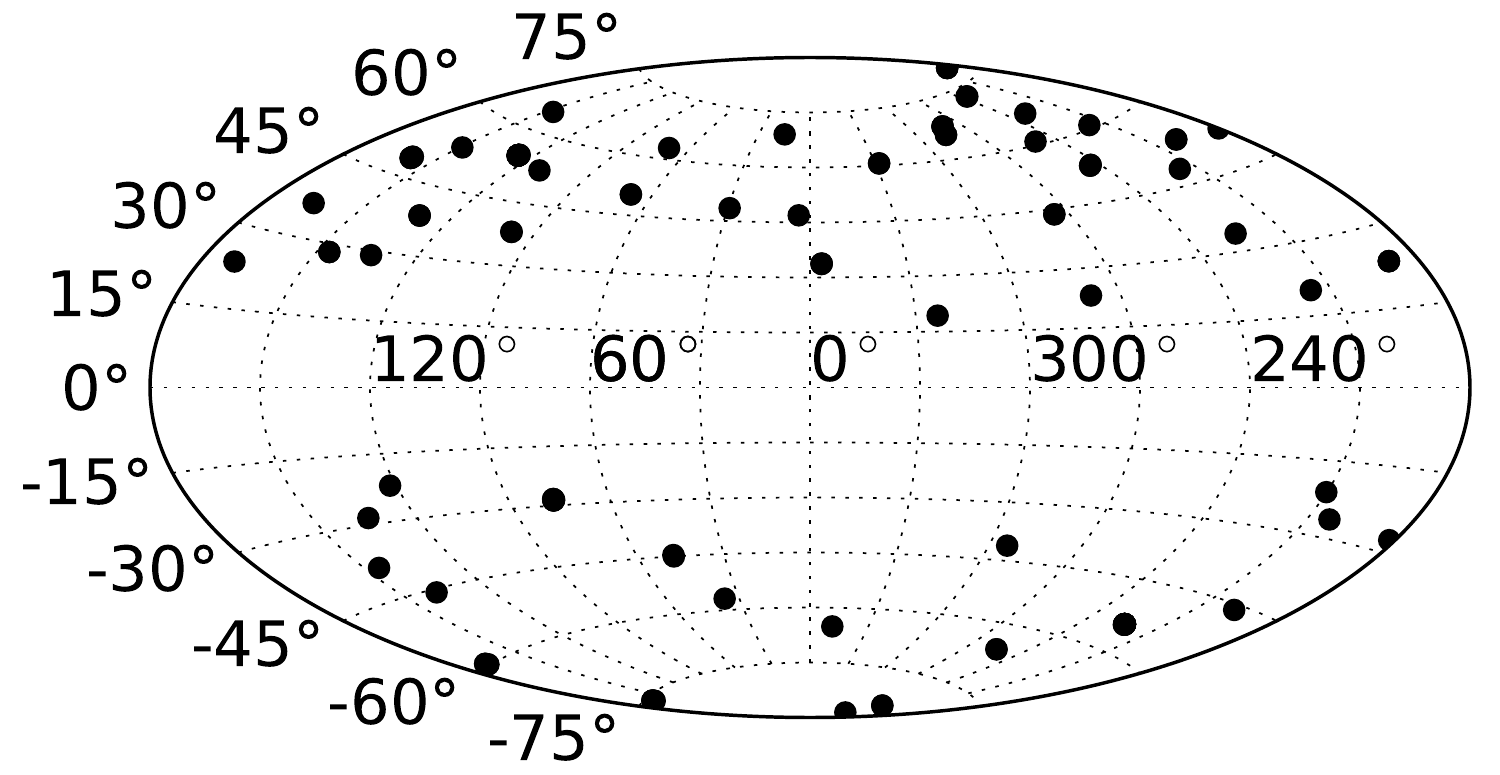}
\caption{Distribution in Galactic coordinates of the 153 ACS grism
  associations included in the release. Many associations are very
  close to each other and thus not discernible
  individually.}\label{fig:sky}
\end{figure}
Whilst around 10,000 spectra have been publicly released from
the GRAPES and PEARS programmes, this represents a small fraction of
the available ACS G800L slitless data. Following on from the HLA NICMOS
slitless archive, and taking account of recent developments in 
the slitless extraction software, a uniform reduction of all the 
ACS G800L slitless data, including GRAPES and PEARS data, presents
strong advantages in terms of a homogeneous and well-characterized survey.
There are also a number of differences between the extraction of ACS
spectra as compared to NICMOS spectra \citep{Freudling08} that are
highlighted in this
paper. Not least for this project was the twenty-fold increase in the
number of extracted spectra. Techniques were developed for assessment
of the content and quality of the released spectra, also employing
machine learning algorithms.
Section 2 describes the steps of data collection and
reduction: basic processing, catalogue generation, wavelength and flux
calibration and quality control. Section 3 describes the empirical
characterization of the extracted spectra by reference to ground-based
spectra, principally in the GOODS fields. Section 4 covers a
description of the released products and some comparisons with
ground-based VLT spectra from the ESO-GOODS survey, and the paper
closes with a summary.  An appendix lists redshifts and line
identifications for selected emission line galaxies with two or more detected
lines.
\begin{table*}[t]
\label{tab:progover}
\centering
\caption{Overview of the provenance of the ACS/G800L data
released in the HLA}
\begin{tabular}{rrrrrl}
\hline
\hline
PI & Proposal ID & $N_{assoc}$& $N_{grism}$& $Exp\mbox{[s]}$&Comment \\
\hline
Perlmutter& 9075&   4&    32  & 24448.0 & \\
Riess   &  9352&   3&    43  & 48790.0 & \cite{Riess2004}; partially processed in GRAPES\\
Fruchter&  9405&   2&    22  & 21990.0 & \\
Yan     &  9468&  43&   377  & 202283.0 & \cite{Drozdovsky}\\
Rhoads  &  9482&  40&   373  & 208607.0 & The APPLES programme\\
Perlmutter& 9727&  2&    44  & 58016.0 & \\
Riess   &  9728&   5&    40  & 43765.0 & \\
Malhotra&  9793&   4&    64  & 92622.0 & The GRAPES programme\\
Thompson&  9803&   2&    36  & 37800.0 & \\
Williger&  9877&   4&    46  & 36560.0 & \\
Ellis   & 10159&   1&    14  & 17300.0 & \\
Riess   & 10189&  14&   109  &104100.0 & \\
Malhotra& 10530&  29&   350  &451781.0 & The PEARS programme\\
\hline
Total:  &      & 153&   1550 &1348062.0& \\
\hline
\end{tabular}
\end{table*}

\section{Image Processing and Target Selection}\label{sec:images}

\subsection{Data preparation}
\label{sec:data_prep}

As a preparatory step, the data from the HST archive were grouped
into so-called ``associations'' before processing them.  An
association is composed of a set of grism and direct images satisfying
a set of rules described below.
The process of building associations was carried out in three
steps: (1) identification of sets of grism images with similar
properties; (2) selection of a direct image for each grism image;
(3) identification of complementary direct imaging in several bands.

For the first process, we defined two grism images as
``compatible'' if \textit{all\/} the following conditions were
fulfilled:
\begin{itemize}
\item HST was using the same guide stars for the grism
  exposures: this ensures very good relative astrometry between the two
  images (cf.\ Sect.~\ref{sec:astr-calib} below);
\item the two grism images have roll angle differences below one
  degree: this ensures that the dispersed spectra overlap when
  the data are stacked and any correction for rotation is small;
\item the images have an overlapping area larger than 50\% of the
  whole field: this ensures that a significant fraction of the objects
  are observed in both exposures justifying the summing of spectra
  to obtain deeper data.
\end{itemize}

Additionally, two grism images not satisfying all criteria above were
still considered as ``compatible'' if there was a chain of compatible
grism images according to the conditions above. This means that two
grism images \emph{A} and \emph{B} were still compatible if there exists
e.g. and image \emph{C} which is compatible with both.
In mathematical terms, we added
a-posteriori a transitivity property to the above relation, which
together with the reflexive and symmetric properties defines an
equivalence relation, that, finally, was used to define an
equivalence class made of compatible grism images.

For each single grism frame, all the direct image frames satisfying the 
same set of criteria described above were identified.  Among these
images, the one showing the smallest offset from each grism image was 
selected to be used in the spectral extraction for the astrometric
calibration of each grism frame.
Finally, we collected all images (in different filters) which had an
overlap of at least 25\% of the ACS field with the union of all grism
frames. This set of images was used to create the reference image of
the association and images in individual bands.

\begin{table}[t]
  \caption{Relevant \texttt{SExtractor} parameters used for the source
extraction on the detection image.}
  \centering
  \begin{tabular}{lcl}
  \hline
    Parameter & Value & Description \\
    \hline
    \texttt{DETECT\_MINAREA} & $5$ & Minimum number of\\
     & & pixels above threshold \\
    \texttt{DETECT\_THRESH} & $3.0$ & Minimum threshold\\
     & & for detection\\
    \texttt{ANALYSIS\_THRESH} & $3.0$ & Minimum threshold\\
     & & for analysis\\
    \texttt{DEBLEND\_NTHRESH} & $8$ & Number of deblending\\
     & & sub-thresholds\\
    \texttt{DEBLEND\_MINCONT} & $0.025$ & Minimum contrast\\
     & & for deblending\\
    \texttt{FILTER\_NAME} & gauss\_2.0\_5x5.conv& Detection filter name\\
    \texttt{SEEING\_FWHM} & $0.105$ & Stellar FWHM $[``]$\\
    \texttt{BACK\_SIZE} & $128$ & Background mesh size\\
    \texttt{BACK\_FILTERSIZE} & $5$ & Background filter size\\
  \hline
  \end{tabular}
  \label{tab:sex}
\end{table}
Since the archive search was restricted to science images
with a full read out on both detectors, it was not necessary to impose a lower
limit for the grism exposure time. As a result of the maximum roll angle difference
within an association, the survey fields from programmes such as GRAPES and
PEARS, that had multiple roll angles for each telescope pointing as part of
their design, were split up into several associations per field.

By following these rules, the grouping of all the ACS G800L grism
pointings resulted in 153 fields almost randomly distributed across
the two Galactic caps. Figure~\ref{fig:sky} shows the distribution of these
pointings and a summary of each association is given in Table A.3. For
each association pertinent details such as the equatorial and Galactic 
coordinates, the orientation on the sky (the orientation of the y-axis
in the detection image)
and the direct imaging filter and exposure times together with the number 
of extracted spectra appearing in the release are listed. A number of 
grism datasets did not meet these criteria to be included in the
final list of associations for processing; they are listed in Table
\ref{tab:failedassocs} together with the reasons for exclusion.

\subsection{Pipeline processing of the data}
The Pipeline for the Hubble Legacy Archive grism data (PHLAG) was
used for the end-to-end data processing of all associations.  The
PHLAG version for processing NICMOS G141 data \citep{Freudling08} was
further developed to optimally deal with ACS data.  The core component
of this pipeline is the spectral extraction software \texttt{aXe}
which was developed by the ST-ECF to reduce the slitless mode data of
HST \citep{Kuemmel09b}.  The data reduction is done in a series of
processing steps which include:

\begin{itemize}
\item data preparation (see Sect.\ \ref{sec:data_prep})
\item data retrieval from the archive (see Sect.\ \ref{sec:pre_processing})
\item removal of cosmic rays on direct imaging using the LACOS algorithm
(see Sect.\ \ref{sec:realignment})
\item re-alignment of the direct images (see Sect.\ \ref{sec:realignment})
\item {\tt MultiDrizzle} of the direct images in the available filters
(see Sect.\ \ref{sec:coadd})
\item object detection with \texttt{SExtractor}
(see Sect.\ \ref{sec:catalog})
\item spectral extraction with \texttt{aXe}
(see Sect.\ \ref{sec:spec_extraction})
\item metadata assembly and generation of the archive data products
(see Sect.\ \ref{sec:metadata})
\item production of the quality control plots 
(see Sect.\ \ref{sec:qualcon})
\end{itemize}

In the following we give a brief description of the various processing steps.

\subsubsection{Pre-processing of ACS data}
\label{sec:pre_processing}
The basis of the data reduction in PHLAG are standard calibrated
images reduced with the {\tt CALACS} pipeline \citep{Hack99}. For slitless
spectroscopic images, {\tt CALACS} applies only a basic calibration
(correction for bias, dark and gain). Direct images are also
flat-fielded in {\tt CALACS}. We have used data reduced with {\tt
  CALACS} version 5.0.4, and for each PHLAG run all images were retrieved
from the ST-ECF HST archive cache \citep{Stoehr09}.

% examples for the previews
\begin{figure*}[t]
\centering
\includegraphics[width=0.45\textwidth]{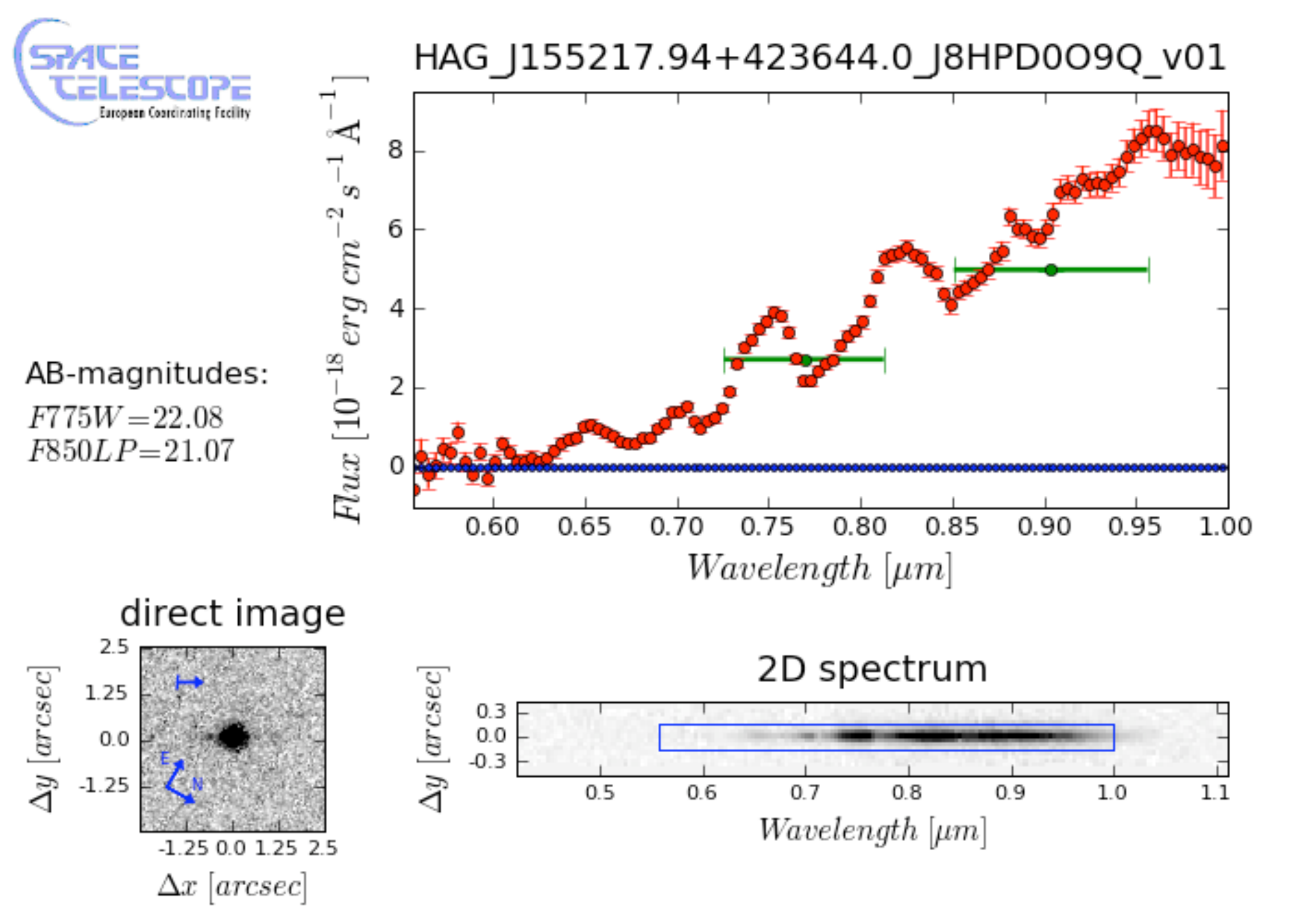}
\hfill
\includegraphics[width=0.45\textwidth]{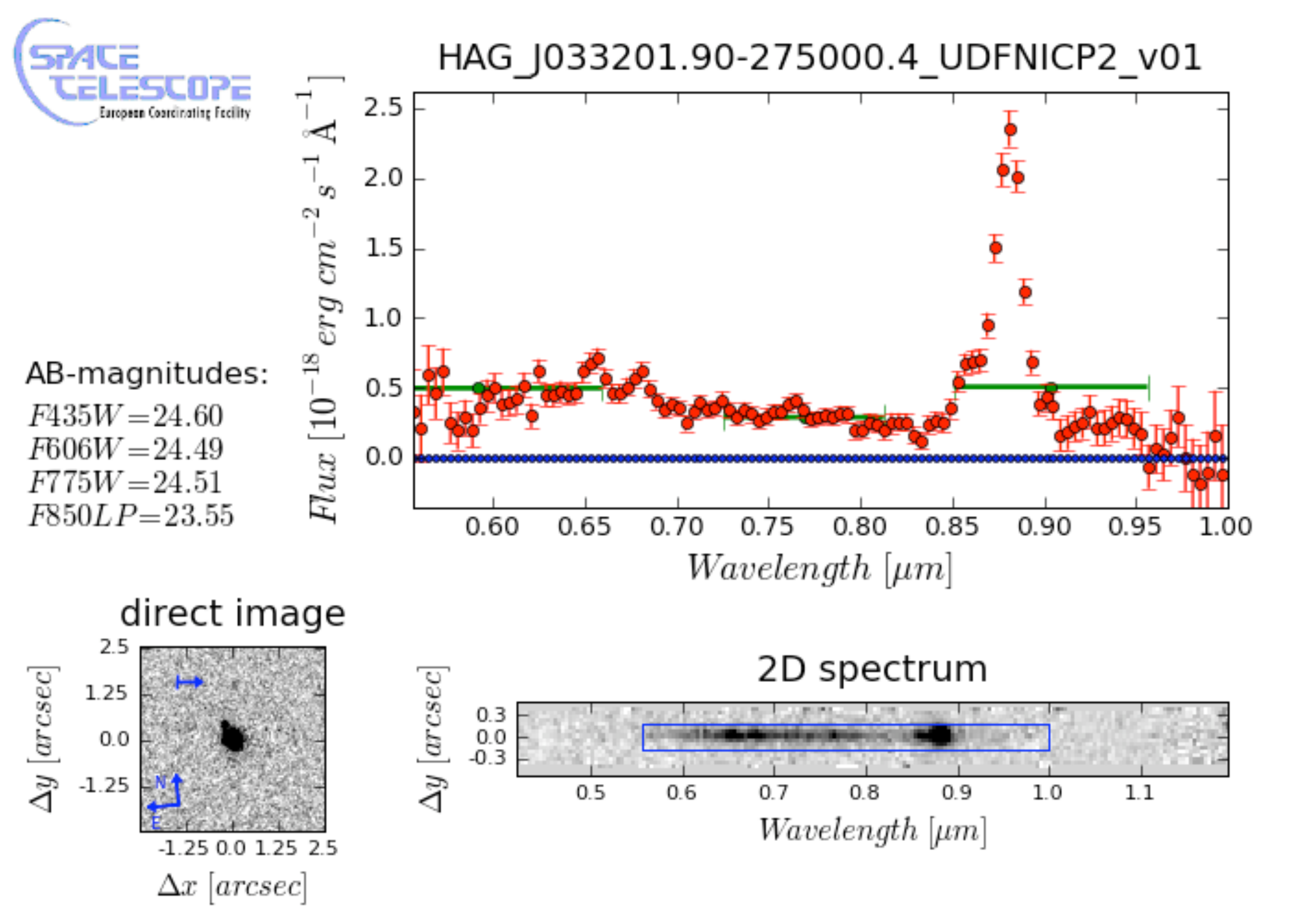}
\\
\includegraphics[width=0.45\textwidth]{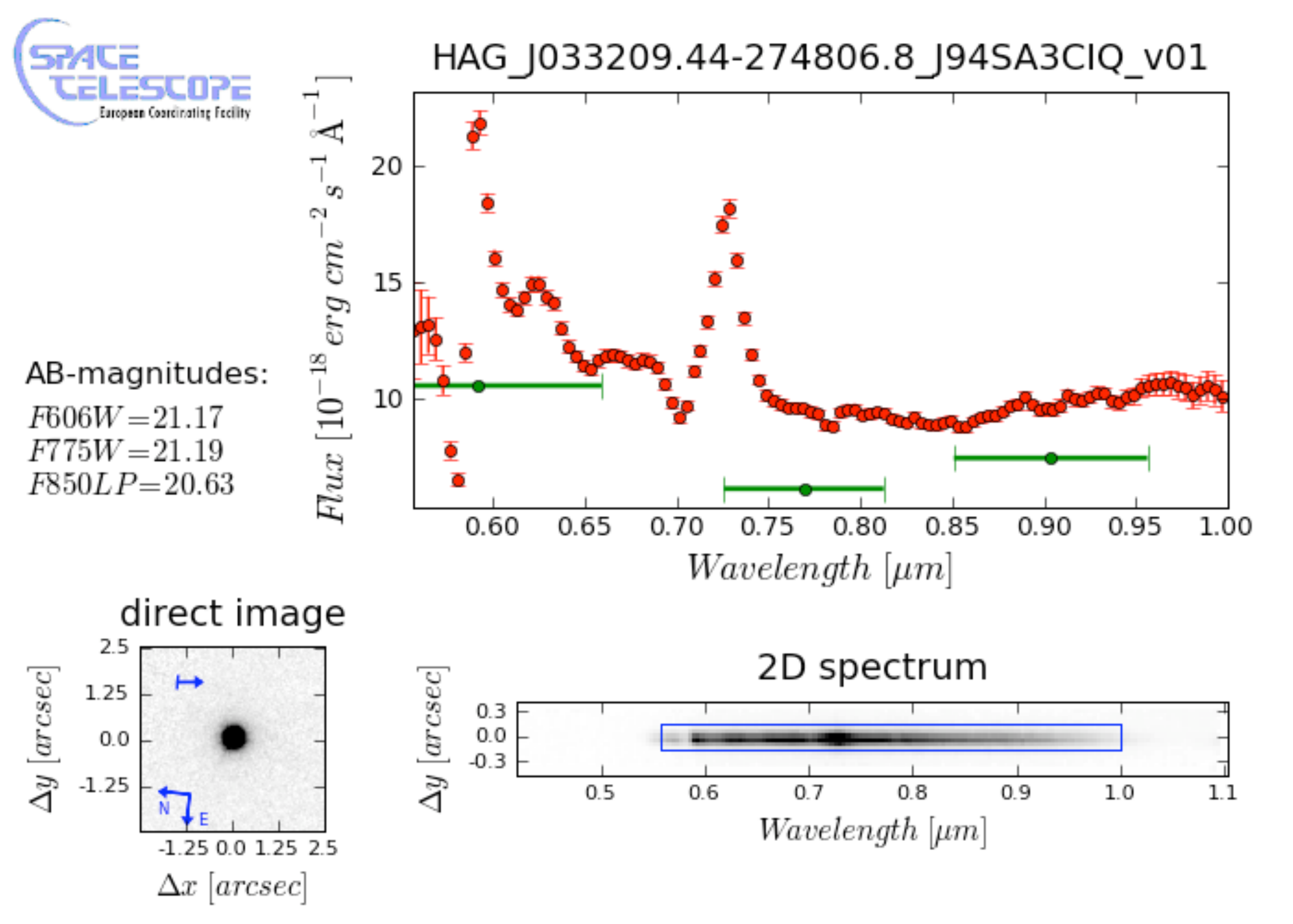}
\hfill
\includegraphics[width=0.45\textwidth]{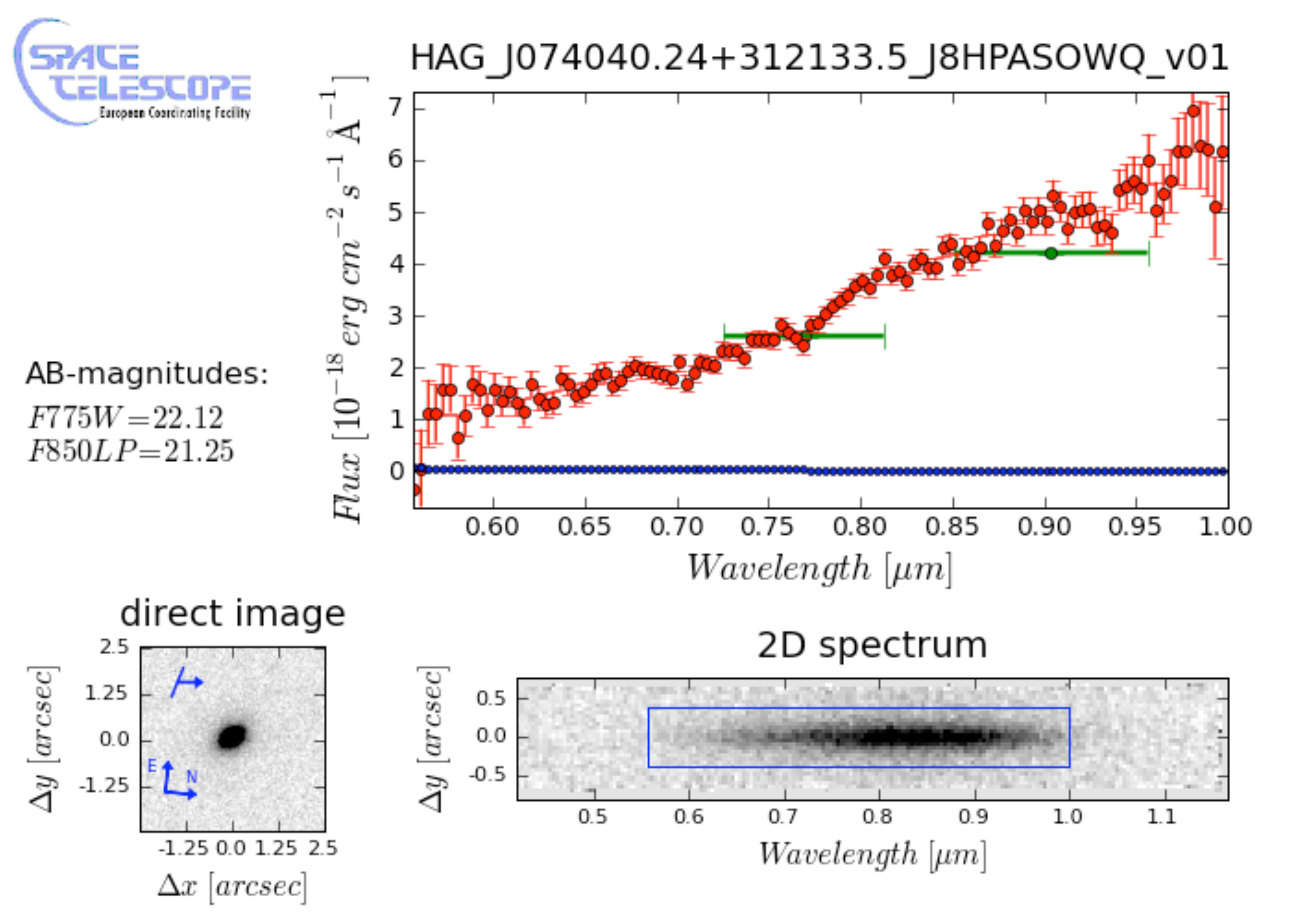}
\caption{Examples of ACS grism spectra as shown in the previews
  included in the release. Blue curves (when present) indicate the
  estimated contamination from near-by sources as a function of
  wavelength. The green data points (horizontal bars) correspond to
  the integrated broad band magnitudes. The blue rectangular box in
  the 2D spectrum shows the effective extracted region where the
  extraction was performed. In the cutout direct image
  (taken from the ``white light'' detection image),
  the ``pseudo-slit'' and the dispersion direction are
  indicated (blue arrow). Top left: an M-star; top right:a low
  redshift star burst galaxy; bottom left: a broad-absorption line QSO
  identified as a Chandra X-ray source at $z=2.78$ \citep{Szokoly04}; bottom right: an early-type galaxy.}
\label{fig:examples}
\end{figure*}
\subsubsection{Re-alignment of the direct images}
\label{sec:realignment}
In order to improve alignment of the direct images, the STSDAS task
{\tt tweakshifts} \citep{koe06} was run on the ensemble of direct
images of each association. {\tt tweakshifts} performs an object
detection algorithm on the ACS images and then determines the residual
shift (no rotation was allowed) by matching the objects on the direct images.
In order to minimize
the number of false detections due to cosmic rays we run the LACOS
algorithm \citep{vDok01} on all direct images prior to {\tt tweakshifts}.
LACOS was run as an PyRAF/IRAF task with the parameters
{\tt sigclip=4.5}, {\tt sigfrac=0.3}, {\tt objlim=3.5} and {\tt niter=4}.
The code of {\tt tweakshifts} was customized to be able to determine
shifts up to $30 \mbox{ pixel}$, in contrast to $\sim 3\mbox{ pixel}$ in the
standard version.

\subsubsection{Co-addition of undispersed images}
\label{sec:coadd}
In each association the direct images were grouped according to the filter 
and then co-added to create a deep image in each filter (with a scale
of $0.05\,\mbox{arcsec/pixel}$). This image combination
is done with the {\tt MultiDrizzle} software \citep{koe06}. We also combine
the dispersed image in order to detect the cosmic ray affected pixels in
the {\tt MultiDrizzle} processing. As in direct imaging {\tt MultiDrizzle}
is able to detect cosmics appearing in individual exposures and does not
extinguish real, compact object signatures such as emission lines and zeroth
order beams that appear in each image.
The co-added grism image from this
operation was not further used in the processing.

\subsubsection{Target Catalogue}
\label{sec:catalog}
The first step in the extraction of slitless spectra is to find
objects on the undispersed image \citep{Kuemmel09b}. Object parameters
required for the extraction are the coordinates, the axis lengths and
the position angle of the major axis.

We used the \texttt{SExtractor} program \citep{Bertin96} to generate the object
catalogues.  For the object detection we generated a deep, ``white
light'' image by co-adding all available deep filter images, using the
exposure time maps as weights.
The photometric information in the filter images is obtained by using
\texttt{SExtractor} in {\it double-image mode} \citep{Bertin08}, whereby one
image (the ``white light'' image) is used for the object detection and
a second image (a filter image) for photometry. For all images the
exposure time maps generated in the co-addition step were used as a
weight map to reduce spurious detections and improve the photometric
accuracy.

The detection list and the photometric lists were combined, and
the resulting catalogue was filtered to remove spurious objects and
bad photometry:
\begin{itemize}
\item we removed spurious objects that were detected but have no photometric
  information;
\item we removed spurious objects that were extremely large ($\mathtt{A\_IMAGE}
  \ge 800 \mbox{ pixel}$) or very large and with a high ellipticity
  ($\mathtt{A\_IMAGE} \ge 100 \mbox{ pixel}$ and
  $\mathtt{A\_IMAGE} / \mathtt{B\_IMAGE} \ge 9.0$);
\item inaccurate photometric measurements
($\mathtt{MAG\_ERR} \ge 2 \mbox{ mag}$) were discarded.
\end{itemize}

The relevant parameters used for the object detection with \texttt{SExtractor}
are given in Table \ref{tab:sex}.

% quality control I
\begin{figure*}[t]
\includegraphics[width=0.45\textwidth]{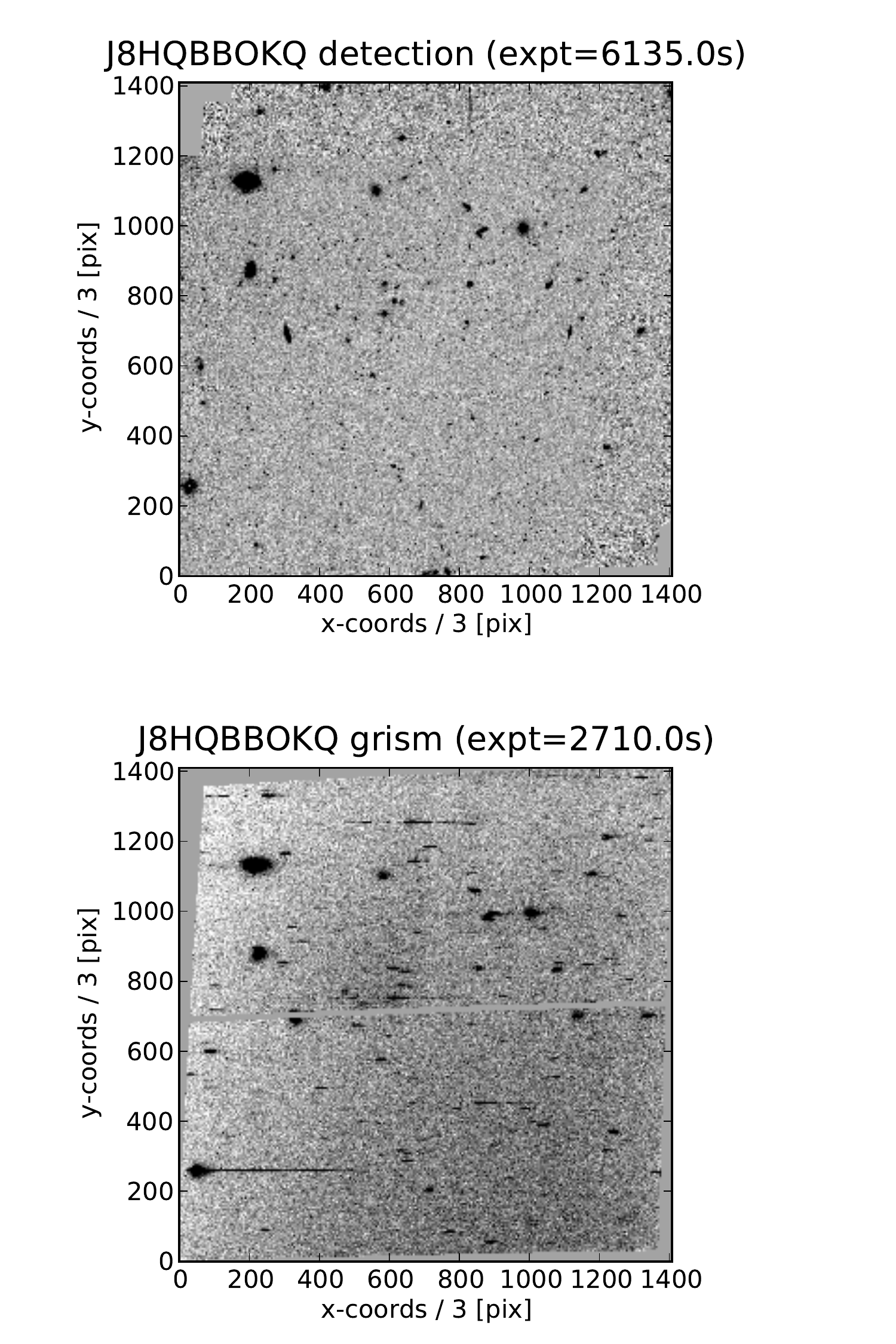}
\includegraphics[width=0.55\textwidth]{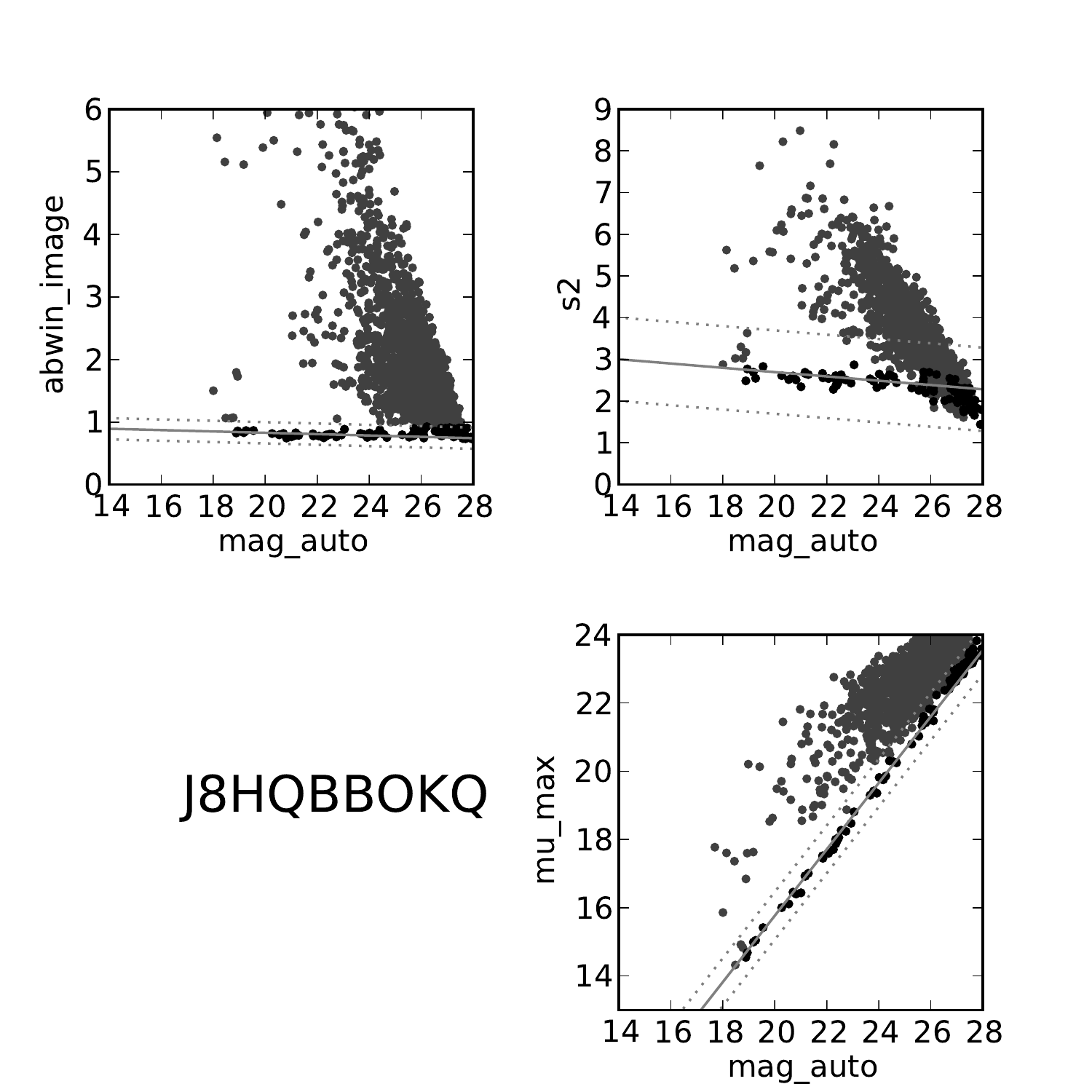}
\caption{Quality control for association \texttt{J8HBBOKQ} I:
the detection image and the co-added grism
image (left side); statistical analysis of the object parameters
from the detection list (right side). The upper left panel shows the object
size, the upper right panel the stellarity parameter {\tt s2}
\citep{Pirzkal2009} and the lower right panel the central surface brightness
as a function of object brightness. The solid and dotted lines mark the exact locus
of point-like objects and the enclosing tolerance area, respectively.}
\label{fig:qualconI}
\end{figure*}
\subsubsection{Spectral extraction}
\label{sec:spec_extraction}
The extraction of spectra from the dispersed images was done with the
\texttt{aXe} spectral extraction software \citep{Kuemmel09b}. The reduction
strategy chosen follows closely the extraction for the NICMOS G141
data as outlined in \cite{Freudling08}. Many of the features developed
for that project have been included into the standard \texttt{aXe} software,
and the ACS grism reduction was done using the version 1.71 of
\texttt{aXe}. Among the processing steps taken, are:
\begin{itemize}
\item the sky background was removed by subtracting a scaled master sky
background image from the science data. The area covered by the spectra of
objects in the master object list was excluded when determining the scale
factor;
\item the extraction direction for each object was chosen to optimize
the spectral resolution as outlined in \cite{Freudling08}
and \cite{Kuemmel10}. Objects with $\sqrt{\mathtt{A\_IMAGE} *
    \mathtt{B\_IMAGE}} \le 1.1\,\mbox{pixel}$ ($=0.055\,\mbox{arcsec}$)
  on the detection image
  were treated as point-like, which
  means $\mathtt{A\_IMAGE} = \mathtt{B\_IMAGE} = 1.1 \mbox{ pixel}$
  and the extraction direction set
perpendicular to the dispersion direction
\footnote{While true point
like sources on ACS/WFC images have lower values
$\mathtt{A\_IMAGE} = \mathtt{B\_IMAGE} \approx 0.95\,
\mbox{pixel}$, slightly extended sources would not profit from an
extraction adjusted for extended objects, hence the choice
of $1.1 \mbox{ pixel}$ as the lower limit.};
\item the contamination of the extracted spectra from neighbouring
  sources was estimated using a model of the dispersed image. The
  latter was generated by adopting the simplest SED for every pixel
  belonging to an object using the fluxes derived from the direct
  images (the so called ``fluxcube'' model, see K\"ummel et al., 2010);
\item the spectra on the individual grism images were re-binned to a
linear scale in wavelength and spatial direction and then co-added to
deep, 2D grism image stamps. The 1D spectra were then extracted from
these co-added image stamps (aXedrizzle, see K\"ummel et al., 2010).
\item extended sources, which here are all objects with a virtual slit
width larger than the 1.1 pix attributed to point like objects (see above),
were flux calibrated using a Gaussian smoothed version of the sensitivity
function \citep{Freudling08, Kuemmel10}.
\end{itemize}

\subsubsection{Aperture correction}
\label{sec:apercorr}
The extraction as described above uses optimized virtual slits that
adjust the slit length to the object morphology. The lower limit for
the object sizes and the chosen scaling parameter in \texttt{aXe} (parameter
\texttt{outfwhm} in the task {\tt axedrizzle}) result in a minimal
aperture with a half-width of $3.3 \mbox{ pixel}$ or $0.165 \mbox{
  arcsec}$. For the minimal aperture the fraction of the total object
flux included in the spectrum can be calculated from the wavelength
dependence of the aperture correction listed in Table \ref{tab:appcorr} 
\citep[from][]{Kuntschner08}.  Using these values, {\it all} extracted spectra
were corrected to the total object flux. Since the aperture correction
in \cite{Kuntschner08} was derived from standard star observations,
this procedure results in the total flux only for point-like objects
and represents only a lower limit for extended sources.
\begin{table}
  \caption{Fraction of the total flux extracted for point-like objects}
  \centering
  \begin{tabular}{rc}
  \hline
    Wavelength [$\AA$] & Value\\
    \hline
 $\le$ 6000.0&  0.892\\
      7000.0&  0.888\\
      8000.0&  0.876\\
      9000.0&  0.844\\
 $\ge$10000.0&  0.679\\
    \hline
  \end{tabular}
  \label{tab:appcorr}
\end{table}

% quality control II
\begin{figure*}[t]
\includegraphics[width=0.5\textwidth]{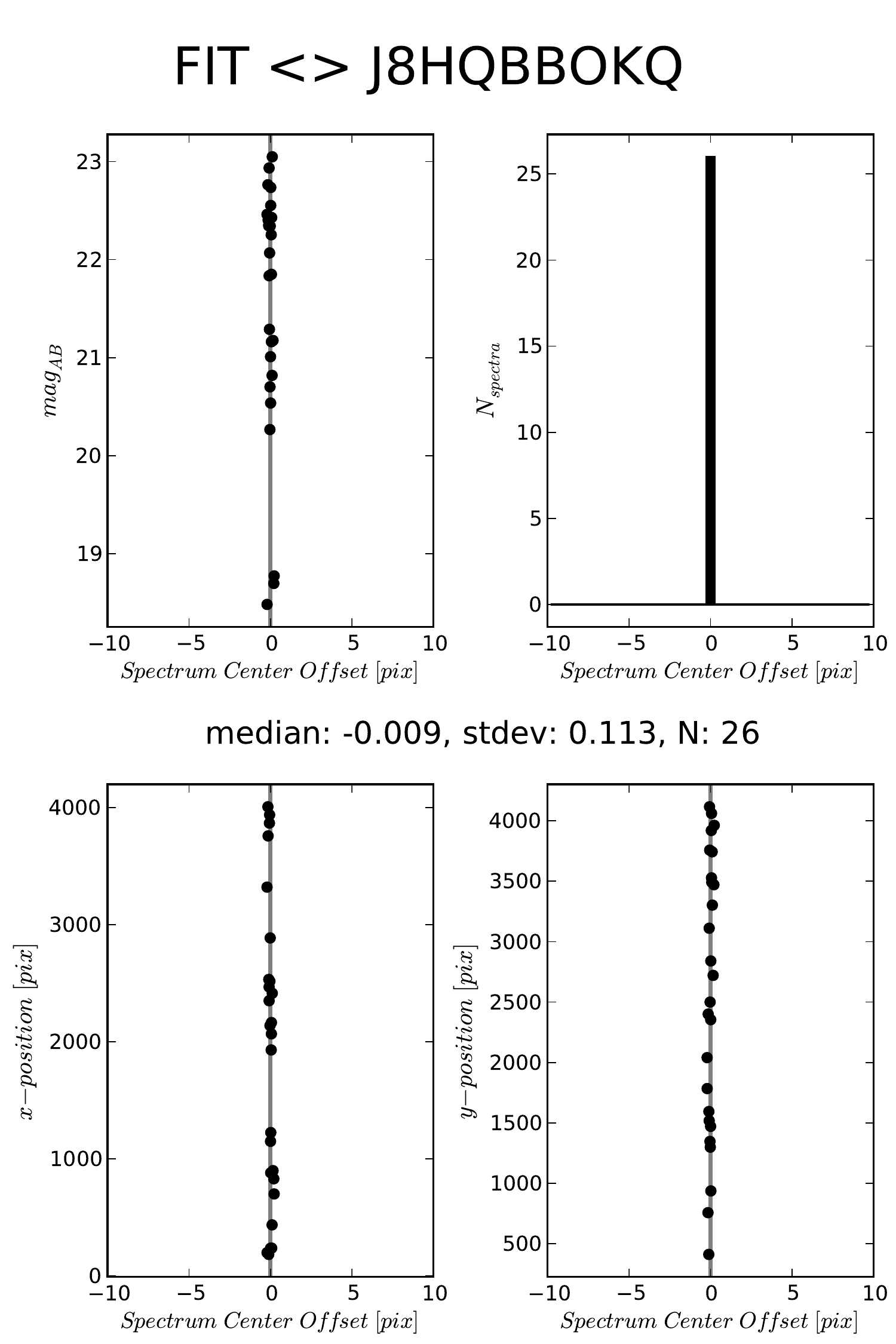}
\includegraphics[width=0.5\textwidth]{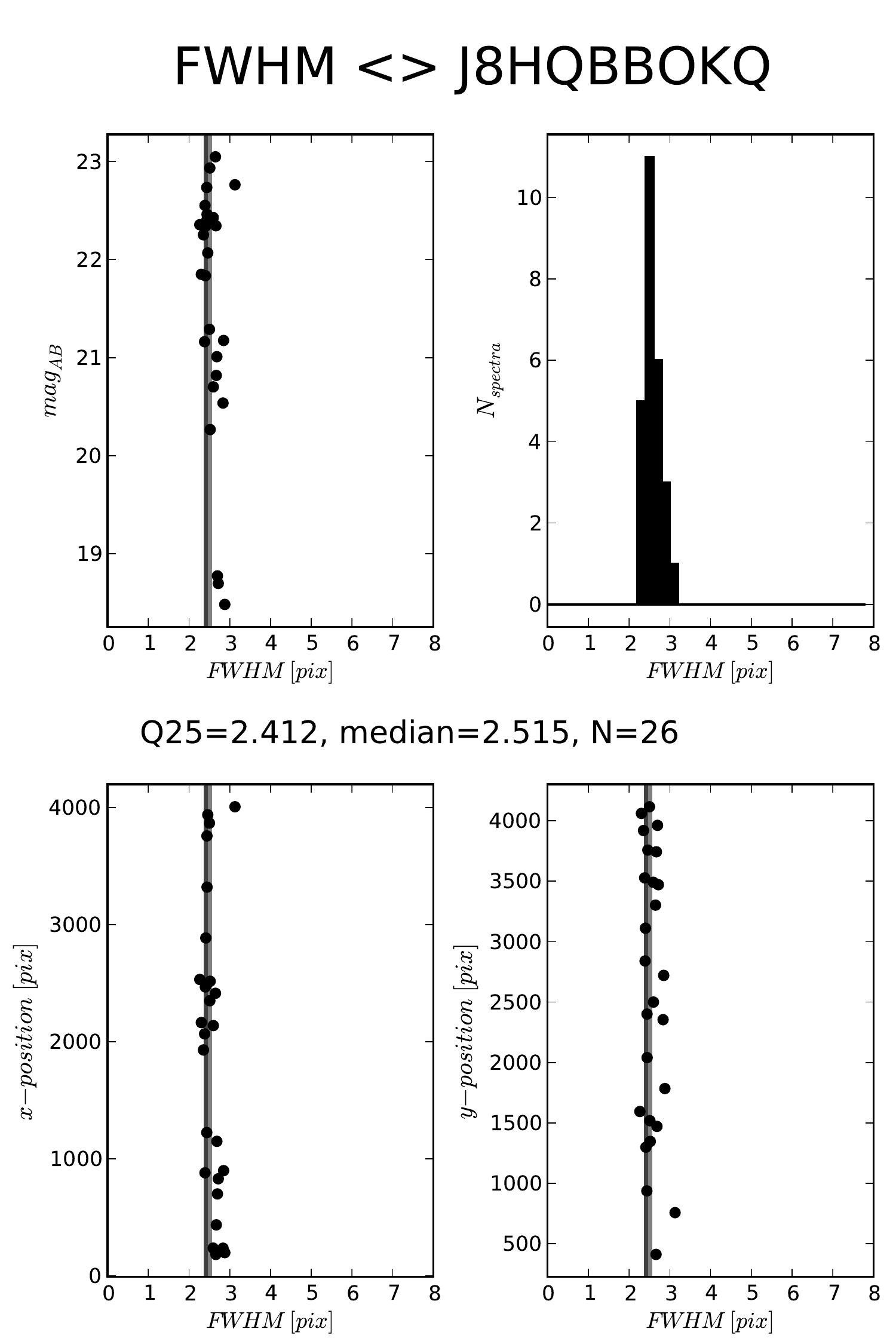}
\caption{Quality control for association \texttt{J8HBBOKQ} II:
Results from fitting Gaussians to the
  spatial profile of the 2D stamp images for bright, compact sources.
  The left side shows the offset of the center from the expected
  position, the right side shows the profile width.}
\label{fig:qualconII}
\end{figure*}
\subsubsection{Generation of archive data products}
\label{sec:metadata}
From the extracted spectra and the co-added filter images, the following data
products are generated to be distributed via the HLA archive:

% quality control III
\begin{figure*}[t]
\includegraphics[width=0.5\textwidth]{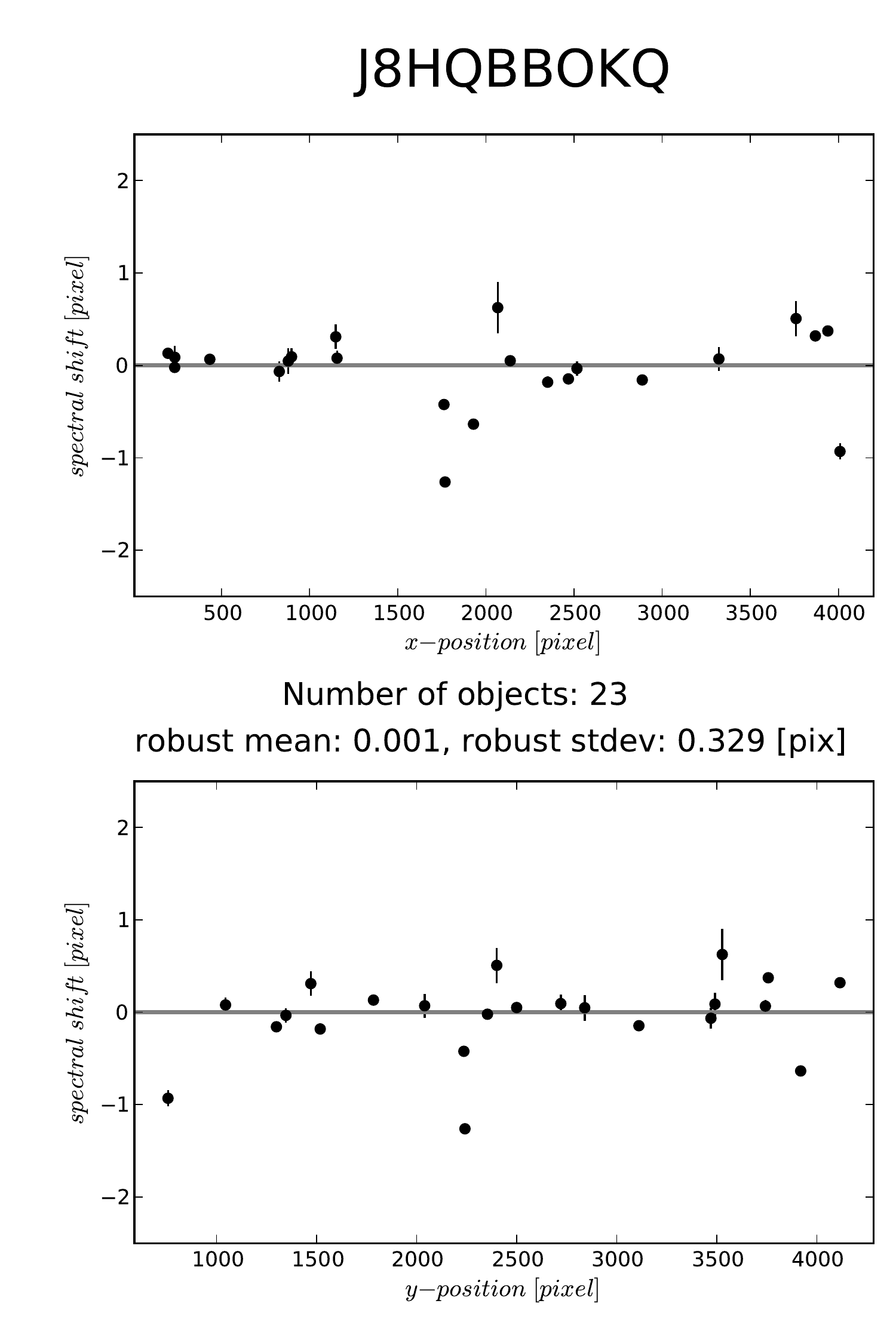}
\includegraphics[width=0.5\textwidth]{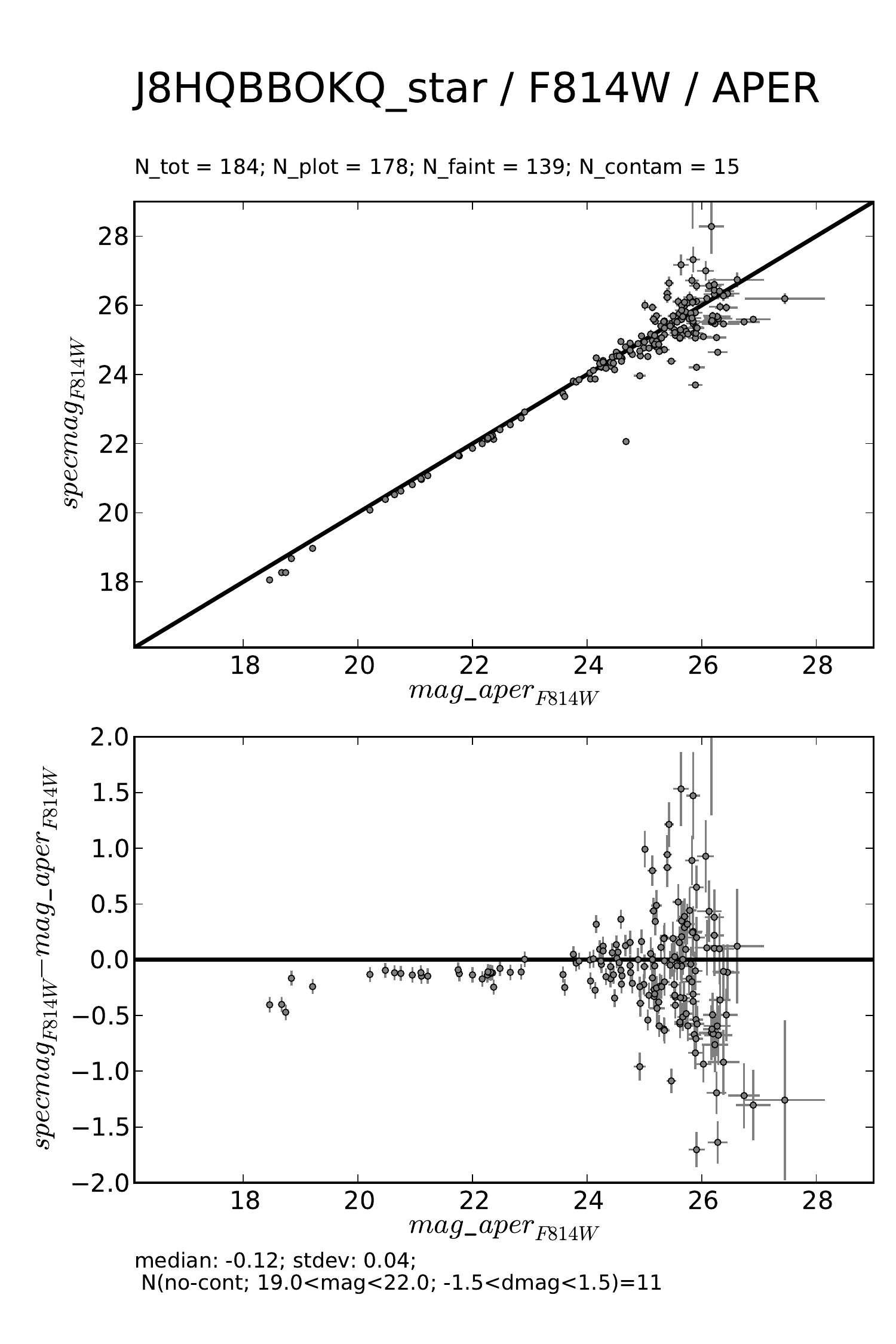}
\caption{Quality control for association \texttt{J8HBBOKQ} III:
The wavelength offset for M-stars
  identified by cross-correlating extracted spectra with M-star
  templates (left side).  A comparison of the direct imaging
  magnitudes and the ``spectral magnitudes'' derived by integrating
  the spectra over the corresponding filter throughput curve (right
  side). The comparison of the direct image magnitudes derived with
  a finite aperture and the ``spectral magnitudes'' that were computed 
  from the aperture corrected spectra (see Sect.\ \ref{sec:apercorr})
  results in an offset from the zero line
}
\label{fig:qualconIII}
\end{figure*}
\begin{itemize}
\item The 1-dimensional spectrum as in FITS table format, following the data
  formatting specified for FITS serialization by the IVOA Spectral
  Data Model version 1.03 \citep{mcd}. Each data point contains the
  wavelength in $\AA$, the count rate in electrons per second and the
  flux expressed in physical units along with associated errors and an
  estimate of the contaminating flux, i.e. the sum of the flux from
  all other sources whose spatial and spectral extent overlaps with
  that of the extent of the extracted source. Additional metadata
  include keywords to describe the contamination, the orientation of
  the dispersion direction on the sky and ``footprints'', i.e. the
  spatial extent on the sky of the extraction region given by the
  height of the extraction aperture and the image size of the source
  (from the \texttt{SExtractor} image parameters) in the dispersion direction.
\item The co-added, 2D grism image as a multi-extension FITS file.
  The various extensions contain the grism stamp image, its error
  array, the exposure time map and the contamination map.
\item The direct image stamp as a multi-extension FITS file. It
  contains the cutout images ($5\mbox{``}\times 5 \mbox{``}$) and the
  corresponding weight images from the detection image and all
  available direct imaging.
\item The preview image such as shown in Figure \ref{fig:examples}.
  The preview combines all basic results in a single image.  It
  contains a spectrum plot with contamination estimate and broad band
  magnitudes, the 2D grism stamp indicating the extraction region and
  the direct image with the orientation on the sky and an illustration
  of the extraction geometry. The extraction length in the
  cross-dispersion direction and the dispersion direction are also
  indicated.
\end{itemize}

The error in the 1-dimensional extracted spectra is based on error
propagation from the error array in the ACS images through the final spectra.
Thus it contains CCD related errors (readout noise, poisson noise) and flux
calibration errors. Errors from flat-fielding and sky background subtraction
are more difficult to quantify on a pixel-to-pixel basis, and have been
taken into account by setting a minimum error threshold of 1\% of the
flux.

The contaminating flux, as estimated from the modelling
of neighouring objects, should help to estimate the level of
contamination in the extracted flux and hence to identify
clean spectral regions for scientific exploitation.

\begin{figure}[t]
\includegraphics[width=\hsize]{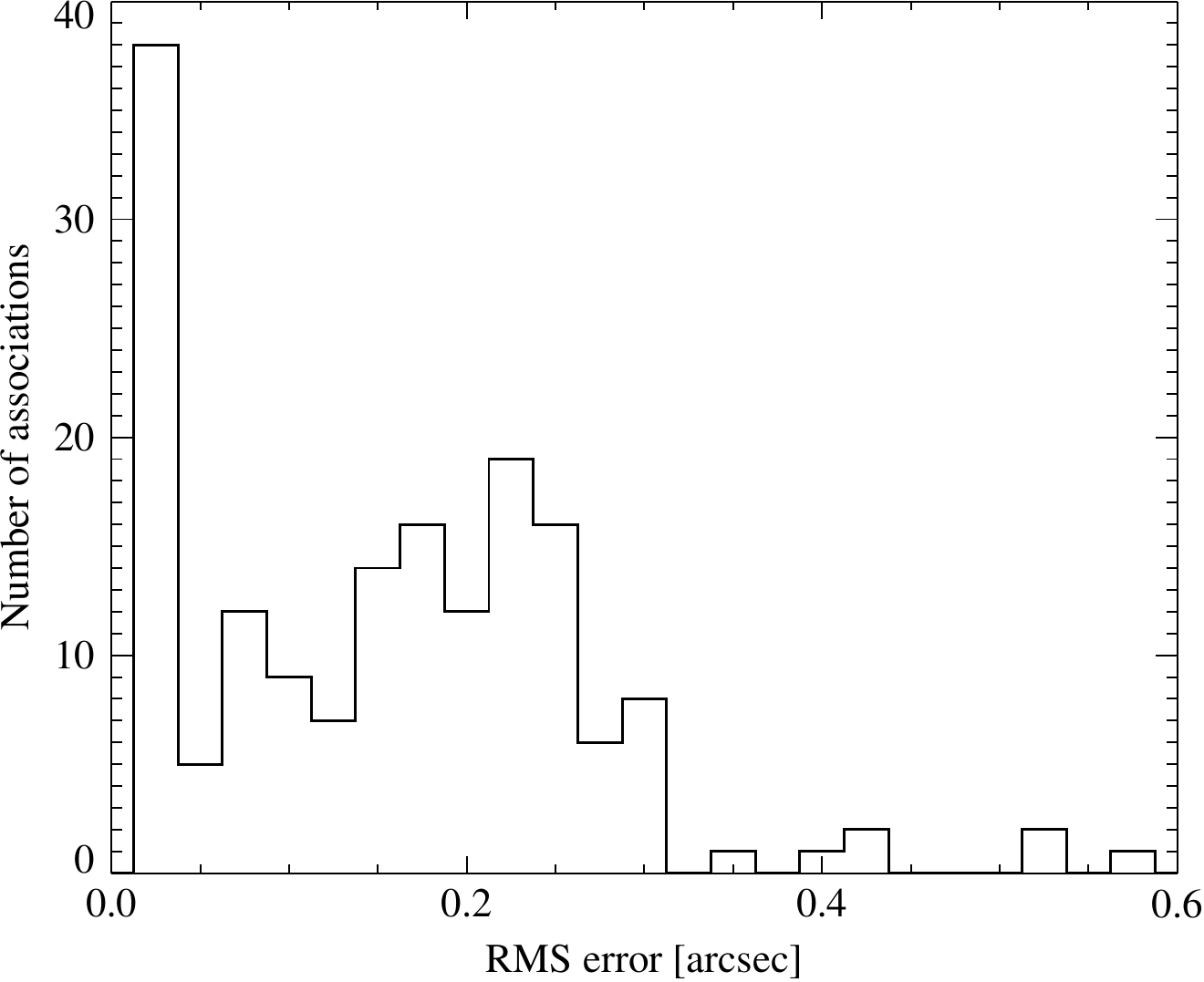}
\caption{Residual astrometric errors after WCS calibration.}\label{fig:21}
\label{fig:WCSres}
\end{figure}
\subsection{Astrometric calibration and accuracy}
\label{sec:astr-calib}
\subsubsection{Quality control of the PHLAG output}
\label{sec:qualcon}
The quality of the PHLAG output is automatically monitored. This is
done by analyzing for each association the products of PHLAG, primarily
the co-added direct images, the source lists and the extracted spectra.
The results are stored as a series of plots representing a particular
aspect. Figures \ref{fig:qualconI}--\ref{fig:qualconIII} show the following
quality control plots produced for the association \texttt{J8HBBOKQ}:

\begin{itemize}
\item A plot of the whole detection image and the co-added grism image
  gives an overview on the field of view and identifies possible
  problems due to e.g.\ stray light (left side of
  Figure~\ref{fig:qualconI}).
\item The plot on the right side of Figure~\ref{fig:qualconI} analyzes
  the size distribution of the detected objects. It summarizes the object
  extent (upper left panel),
  the stellarity parameter \texttt{s2} (upper right panel; the ratio of
  the flux within a circular
  aperture of radius 1 pixel to the total isophotal flux as measured by
  \texttt{SExtractor}, see Pirzkal et al., 2009) and the central surface
  brightness as a function of source brightness lower right panel.
  The locus of
  point-like objects is marked with the dotted parallel lines, and problems in
  the re-alignment of direct images result in an offset from this
  locus, and were readily apparent. The quality flag attributed to each
  association in Tab.\ A.3 is mainly based on this plot.
\item For each 2D grism stamp image the cross-dispersion profile was extracted 
and fitted by a number of Gaussians at different wavelengths.
 Possible offsets of the
  spectrum centres from their nominal position (Figure
  \ref{fig:qualconII}, left side) or large widths (Figure
  \ref{fig:qualconII}, right side) point to alignment problems between
  the slitless images and their associated direct images.
\item M-stars were identified by cross-correlating the 1D-spectra of
  point-like objects (from the \texttt{SExtractor} catalogue) with a set of
  M-star templates taken from the BPGS atlas \citep{Gunn83}. Since the
  zeropoint of the wavelength calibration depends on an accurate
  translation of the object position from the direct to the slitless
  images, a wavelength offset (see Figure \ref{fig:qualconIII}, left
  side) identified in the M-stars would point to alignment
  problems between the grism images and their associated direct
  images.
\item The 1D-spectra are integrated over all direct imaging filters
  (if the filter passband was covered by the wavelength range of the
  spectra). These ``spectral magnitudes'' are then compared to an
  aperture magnitude derived from the filter images. The right side of
  Figure \ref{fig:qualconIII} shows such a comparison for compact
  objects of one association and the \texttt{F814W} filter. While
  brightness differences of $\approx 0.15 \mbox{ mag}$ as shown in
  Fig.~\ref{fig:qualconIII} are expected due to the usage of aperture
  corrected spectra but finite photometric apertures, larger
  differences or even systematic trends  identify problems
  in either the direct imaging or the spectra extraction.
\end{itemize}

\begin{figure}[t]
\includegraphics[width=\hsize]{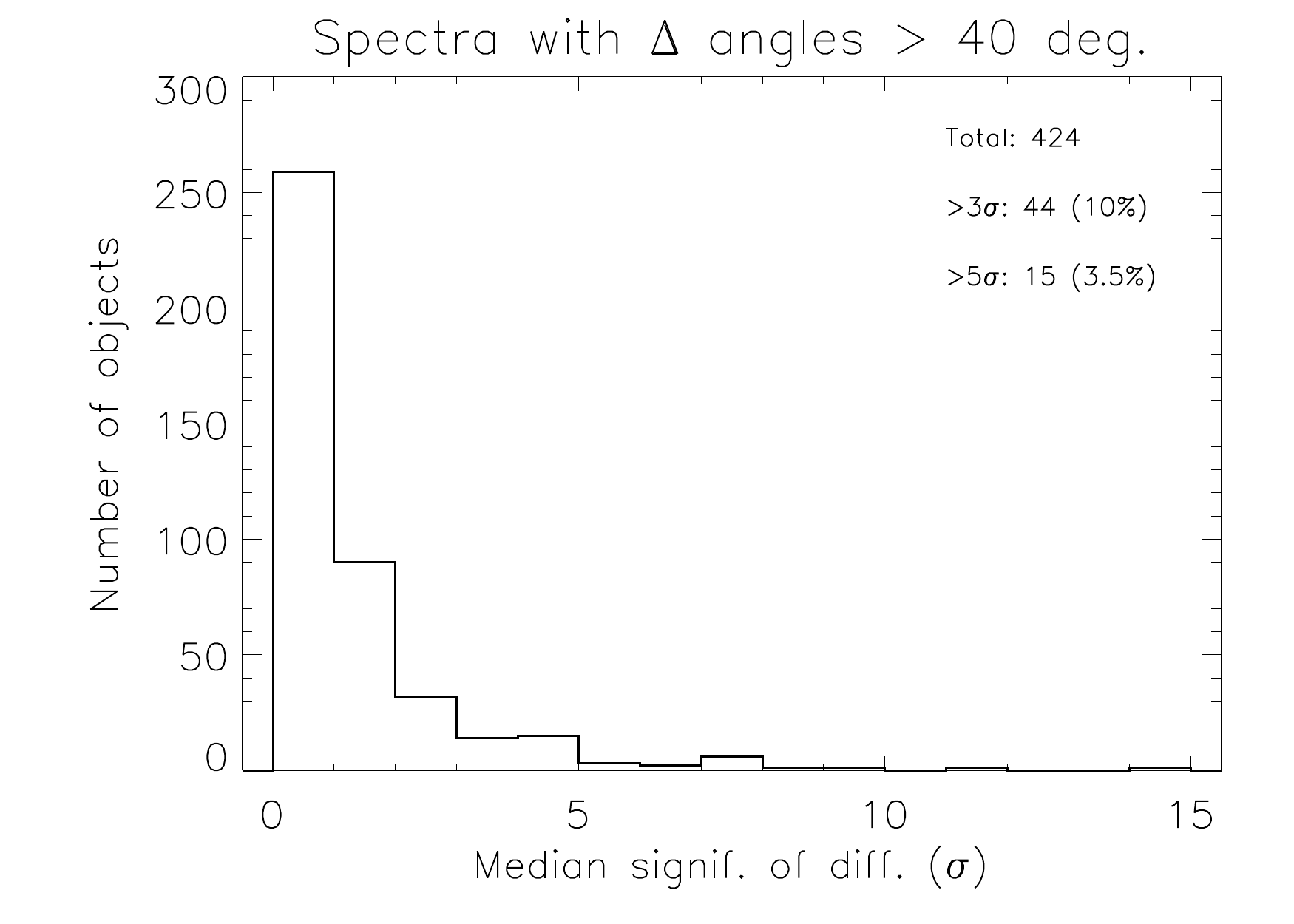}
\caption{Median $\sigma$ of all spectral bins of each pair of
  spectra, where both spectra have low contamination, and the roll
  angle between the two measurements is $>40^{\circ}$.}
\label{fig:medsig}
\end{figure}
On the basis of these plots problems were identified on an association
by association basis and corrected prior to production of the final
release data.  As a result of the quality control, the following
rules were defined in order to ensure homogeneous quality for the
products of all associations:
\begin{itemize}
\item a minimum of 2 grism images in an association is required;
\item a minimum of 2 direct images in one filter is required for an
  association;
\item direct images that could not be re-aligned with the \texttt{tweakshift}
tasks (see Sect.\ \ref{sec:realignment}) were discarded.
\item the direct imaging in the filters blue-wards of the slitless
  spectrum range (\texttt{F435W}, \texttt{F475W} and \texttt{F555W}) are
  discarded if the re-alignment or the co-adding failed;
\item associations with an object density higher than $2,000$ per square
  arcminute are discarded, since in this case overlap of spectra is
  such that the majority of spectra are multiply contaminated and there is
  insufficient background region available for background estimation;
\item the accuracy of the background subtraction for the grism data
  was monitored. Grism images with large residuals, e.g. due to
stray light from a bright star or a low surface brightness contribution
from a large galaxy, were discarded. As a result, some associations taken close
to M31 could not be reduced (see Table \ref{tab:failedassocs}).
\item significant offsets either in the spatial direction or in the
  wavelength direction as identified in Figs.\ \ref{fig:qualconII}
and \ref{fig:qualconIII} were
  translated into shifts of the object position and applied. E.g.\ for
  the association {\tt J8HQBBOKQ} the offsets $-0.45 \mbox{ pixel}$
  and $+0.25 \mbox{ pixel}$ had been applied in x- (spectral-) and
  y- (spatial-) direction, respectively. In general, shifts in x-direction
were applied if at least 5 M-stars could be identified for an association and
the shift was greater than 0.25 pixel with 2-$\sigma$ significance.
Shifts in the y-direction were applied if larger than 0.25 pixel.
\end{itemize}

\begin{table*}
\caption[]{Reference catalogues used for astrometric calibration.}\label{tab:refcats}
\centering
\begin{tabular}{lccc}
  \hline
  Catalogue & Acronym & Reference & Accuracy \\
  \hline
  USNO CCD Astrograph Catalog& UCAC2 & \cite{2004AJ....127.3043Z} & 0.06
  arcsec\\
  Two Micron All Sky Survey& 2MASS  &\cite{2006AJ....131.1163S} & 0.10 arcsec\\
  Sloan Digital Sky Survey& SDSS-DR5  &\cite{2007ApJS..172..634A} & 0.15
  arcsec \\
  USNO-B1.0 catalog & USN0& \cite{2003AJ....125..984M} & 0.20 arcsec\\
  Guide Star Catalog II & GSC2.3.2 & \cite{gsc2} & 0.30 arcsec\\
  \hline
\end{tabular}
  \label{tab:astrocats}
\end{table*} 
As a result of this first quality control process, 15 associations
were discarded and hence not included in this release, leaving the
total of 153 associations shown in Fig.\ \ref{fig:sky}
(see Tables \ref{tab:failedassocs} and A.3).

The astrometric coordinate system of the raw ACS images is specified
in the WCS keywords in the image headers.  The accuracy is ultimately
limited by the accuracy of the guide catalogue used for pointing the
telescope.  For earlier data, the Guide Star Catalog~I
\citep{1990AJ.....99.2019L} was used, and since 2000 the Guide Star
Catalog II \citep{gsc2} is used.  
The absolute accuracy reaches $0.3 \mbox{\ arcsec}$ over a
large fraction of the sky, but errors can be as high as several
arcseconds towards the edges of the scanned plates. Additionally,
misidentification of sources or confusion effects can, in some cases,
create inaccuracies in the raw ACS astrometry as large as $10 \mbox{\ arcsec}$.

In order to improve the astrometric accuracy of the ACS direct images,
and thus of the sources associated to released spectra, we
cross-correlated source catalogues extracted from the ACS fields with
several reference astrometric catalogues: UCAC2, 2MASS, SDSS-DR6,
USNO-B1, GSC2.3.2, and the GOODS WFI R-band catalogue where relevant (see
Table~\ref{tab:astrocats}).  When multiple matches were found, the
most accurate of these catalogue was used.  The fitting procedure
allows for simple shifts of the astrometry, but not for rotation which
we found to be generally negligible in HST observations.  We used
several techniques to avoid false matches, including sigma-clipping
and cluster analysis; we also introduced magnitude cuts for each
astrometric catalog (so that, for example, a 2MASS star would not
match an object of R mag. of 22).

We evaluated the astrometric accuracy of each association
from the standard dispersion of all offsets measured between objects found in
the association and the corresponding matched reference objects from
the astrometric catalogs. While the residual errors are equally distributed
in {\tt RA} and {\tt DEC}, Figure \ref{fig:WCSres} shows the histogram
of its absolute value for all associations after applying the WCS calibration.
The error depends strongly on the
number and quality of the matches obtained, but in any case should
provide a good upper limit to the actual accuracy of the astrometric
solution. As can be seen in Fig.\ \ref{fig:WCSres},
the accuracy for the large majority of the
fields is below $0.2 \mbox{ arcsec}$ (with median error $0.15 \mbox{arcsec}$).
This accuracy is confirmed by a check in the GOODS fields where accurate
($50 \mbox{mas}$) astrometry is available.

\subsection{Wavelength calibration}
\label{sec:wl-calib}
In-orbit wavelength calibration for G800L spectra was established by
observations of the Wolf-Rayet star WR96 and verified with
observations of the planetary nebula LMC-SMP-81 \citep[programmes 9568
and 10058, see ][]{Larsen05}.  The observations of the Wolf-Rayet star cover
20 independent locations across chip\,1 and chip\,2 while LMC-SMP-81
was observed at three different positions. The wavelength calibrations
were carried out with a direct least-squares fit of the G800L spectra
to a (smoothed) spectrum of WR96 obtained from the ground
\citep{Pasquali06}. This approach has the advantage that blends are
automatically accounted for, and thus does not rely on the ability to
accurately measure the wavelengths of individual spectral features
\citep{Larsen05}.
With the verification of the wavelength zero point via M-star
spectra (see Sect.\ \ref{sec:qualcon}) the  overall absolute  wavelength
calibration error is estimated to be smaller than 1 pixel, i.e. $40\mbox{\AA}$.

\subsection{Flux calibration of 1D spectra}
\label{sec:flux-calib}
The flux calibration for the G800L grism was derived from observations
of the primary HST standard star G191B2B at 11 different positions
covering both chips of ACS and cross-checked with observations of the
standard star GD153 \citep[programmes 9029, 9568 and 10374, see][]{Kuntschner08}. The absolute flux calibration of the 1$^{st}$
order G800L spectra for point sources is accurate to better than 2\%
for the spatial positions covered by the standard stars and
wavelengths from 6000 to 9500~\AA. Including errors from the large
scale flat-field, the absolute calibration is better than 5\% for point
sources.

\subsection{Duplicate spectra} 
Several grism pointings overlap, in all areas of the sky, and in
particular in the GOODS-S field resulting from the GOODS and PEARS
programmes. This offers an opportunity to test the robustness of the
reduction and extraction routines. We thus searched the complete
catalogue of released spectra, including $47,919$ spectra in 153
associations, for companion spectra within a five arcsecond radius. We
then plotted the radial distribution of the companion spectra, and
found a clear break at $\sim 0.3\arcsec$. We choose this as the radius
within true matches are to be found. In the total sample, $32,149$
objects have unique spectra and $8,185$ have duplicate spectra,
ranging from a single pair of duplicates, to up to 13 spectra of the
same object. $\sim 40$\% of all the duplicate spectra are in the
GOODS-S field and $\sim 20$\% in the HUDF.

For the test of the accuracy of the data reduction and extraction,
duplicate spectra with contamination $< 20$\% in the whole spectral
range were selected.  Further, to test the most extreme cases, only
duplicates where the roll angle differences between observations was
$> 40^{\circ}$ were compared. No constraint was put on the size of the object. 
The significance on the agreement between
the two spectra was calculated as the median significance ($\sigma$)
of the difference between the two spectra, weighted with the error on
each spectrum, over all wavelength points. The resulting histogram of
median $\sigma$ is shown in Fig.~\ref{fig:medsig}.

Of the spectra measured in this way, 61\% had median significances
within $1\sigma$, and $10$\% had values $>3\sigma$. The most common
reason for a large $\sigma$ are extremely small error-bars on very
bright objects. Overall, the agreement between spectra, even with a
large roll angle between the observations, is remarkably good.

\subsection{Quality control of extracted spectra}
\label{sec:spec_QC}
In the pre-release of the ACS/G800L data set in May 2009, which was
restricted to two field in the GOODS south area \citep{Kuemmel09c},
the number of extracted spectra was small enough to allow an individual,
visual inspection. During this interactive quality control process ~40\% of
the extracted spectra were discarded, which led to the final release
of $1,235$ spectra. The main reasons for the rejection was
a high contamination from nearby sources, but also the presence of deviant
pixels that are either saturated or lying near the
edges of the observed fields.

\begin{figure}[t]
\includegraphics[width=0.9\hsize]{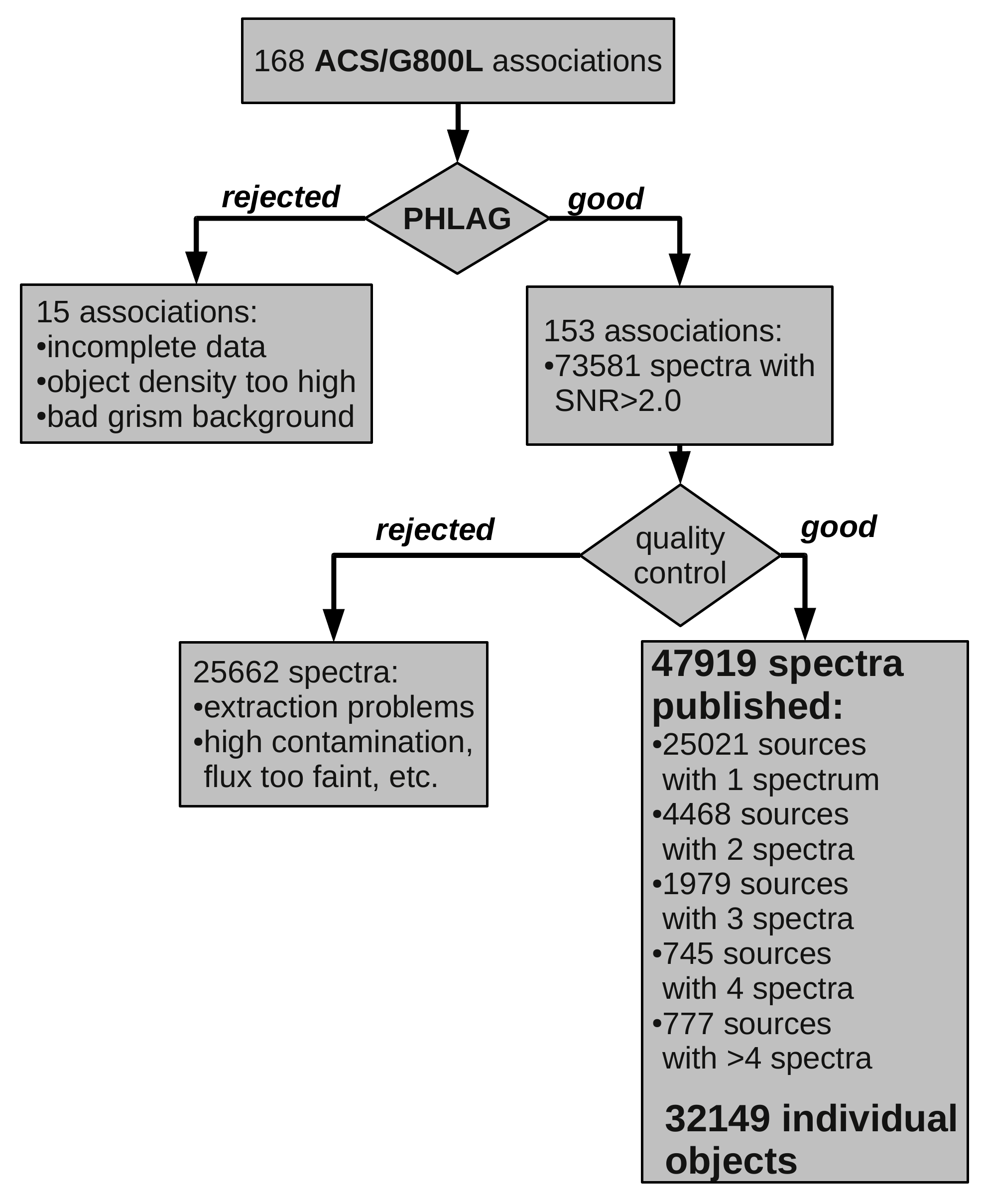}
\caption{Flow diagram of the data processing and quality control, leading to the released sample of HLA/ACS spectra and related number of individual unique sources. }\label{fig:flchart}
\end{figure}
A visual inspection of the full sample of $73,581$ spectra produced in
this release was impossible to do on reasonable time-scales. We
therefore have explored ways of automatic classification using
machine-learning techniques. The idea is to first train an algorithm
with a relatively small number of hand-classified spectra using
measured values such as the estimated contamination fraction, the
signal-to-noise ratio, the magnitude, the position of the maximum of
light on the 2D spectrum or the exposure time. In a second step then,
ideally the remaining spectra can be classified automatically with the
algorithm returning the classification of the spectrum at hand by
finding the classification of the most similar spectrum in the
training set.

Three of us (JW, HK and PR), acted as quality control scientists, inspecting
$2,020$ ($\sim 3\%$ of the total sample) randomly
selected spectra from the full sample and classified them independently
into "good" and "bad". The $241$ spectra with mismatching classifications
were re-examined until a consolidated classification by all three scientists
was achieved.
This process resulted in a very clean and homogeneously training sample
for the automatic classification, which is a key to get high automatic
classification success rates.

The automatic classification was then carried out using the \texttt{weka}
software package \citep{Hall09}. The machine learning algorithms were
trained with $2/3$ of the consolidated training set and the quality of
the classification was measured by applying the trained algorithm on
the independent remaining $1/3$ of the inspected set. Several algorithms
showed similar performance for the task at hand. The final algorithm
chosen was \texttt{ClassificationViaRegression} using an M5P tree, which
resulted in a very high total classification rate (90\%).
Also the rate of false negatives ("good" spectra classified as "bad") was
only $2.5\%$ and thus very low with this algorithm.
This is important as it turned out that there were a surprisingly
large percentage of spectra in the sample that are borderline cases
(roughly $15\%$) between the spectra that are unambiguously publishable
and those that are not. With the choice of this algorithm we opted to
include the spectra into the "good" sample if in doubt. The
classification quality of $90\%$ achieved by the algorithm is comparable
or even slightly better than the quality that each of the quality control
scientist achieved individually when inspecting the training sample.

The classification algorithm was then run on the $97\%$ of the remaining 
spectra of the full sample. The $21,416$ spectra falling in the "bad" 
category were discarded. The $50,145$ spectra from the "good" category were 
passed on to post-classification analysis.

During the training of the algorithm it became apparent that there are 
classes of spectra that are clearly flawed but are classified by the 
algorithm as ``good''. The reason for this is that the number of such cases 
in the training sample is too small for the algorithm to build up 
expertise. We call such spectra "catastrophic failures". 

One such class of spectra is characterized by saturated centers (67 spectra).
A second class of $1,628$ spectra displays dramatically raising SED's at
both ends of the spectra range, which is caused by an underestimated
object size and thus a too narrow smoothing kernel in the flux calibration
(see Sect.\ \ref{sec:spec_extraction}).
These classes of spectra could be identified via specific detection
algorithms, and their small number allowed a visual inspection
and selection of all members.

\begin{figure}[t]
\includegraphics[width=0.9\hsize]{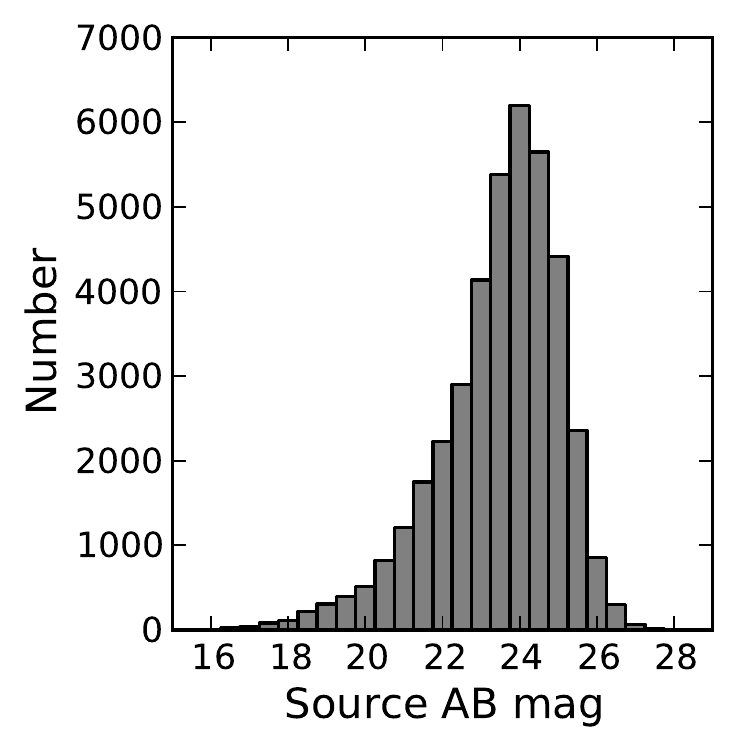}
\caption{Magnitude distribution of the $32,149$ individual sources
  associated to the HLA/ACS spectra. AB F775W magnitudes are used for
  most of the objects.}\label{fig:maghisto}
\end{figure}

The remaining set of "good" spectra was then looked-at by
concatenating 100 previews into a poster-like web interface for rapid
inspection. 1951 spectra (i.e. 4\% of the remaining sample) were
discarded in this process resulting in a total of $47,919$ "good"
spectra. The effort for skimming through these spectra was roughly
equivalent to that of carefully classifying the training set. The
machine classification therefore saved more than 90\% of the time that
would have had to be spent on a complete visual classification. The
entire data processing and quality control process of associations and
spectra which lead to the final released sample of ACS HLA spectra is
summarized in the flow diagram shown in Fig.~\ref{fig:flchart}. The
magnitude distribution of the $32,149$ unique sources is shown in
Fig.~\ref{fig:maghisto}.

\subsection{Comparison of Spectra with VLT Data}
Twenty-nine grism pointings are located within the GOODS-S field
\citep{Giava04}, where many reduced spectra from other
surveys are publicly
available. This allows the comparison of our spectra with ground
based spectra of the same objects as an external quality control.
The catalogues of spectra used in this test come from the
ESO/VLT FORS spectrograph \citep{Szokoly04, Vanzellla08}
and from the ESO/VLT VIMOS instrument \citep{LeFevre04, Balestra2010}.

\begin{figure}
\includegraphics[width=\hsize]{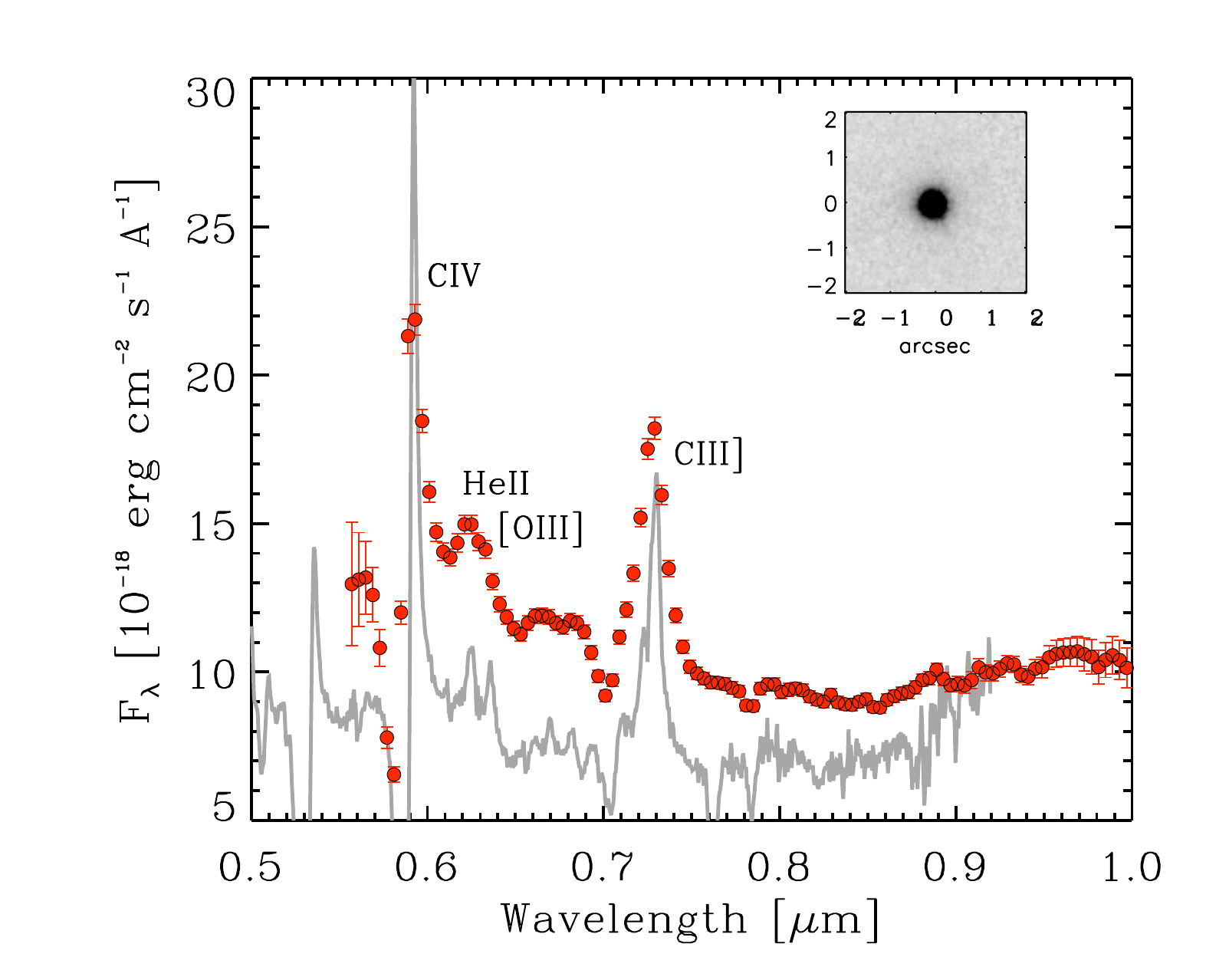}\\
\includegraphics[width=\hsize]{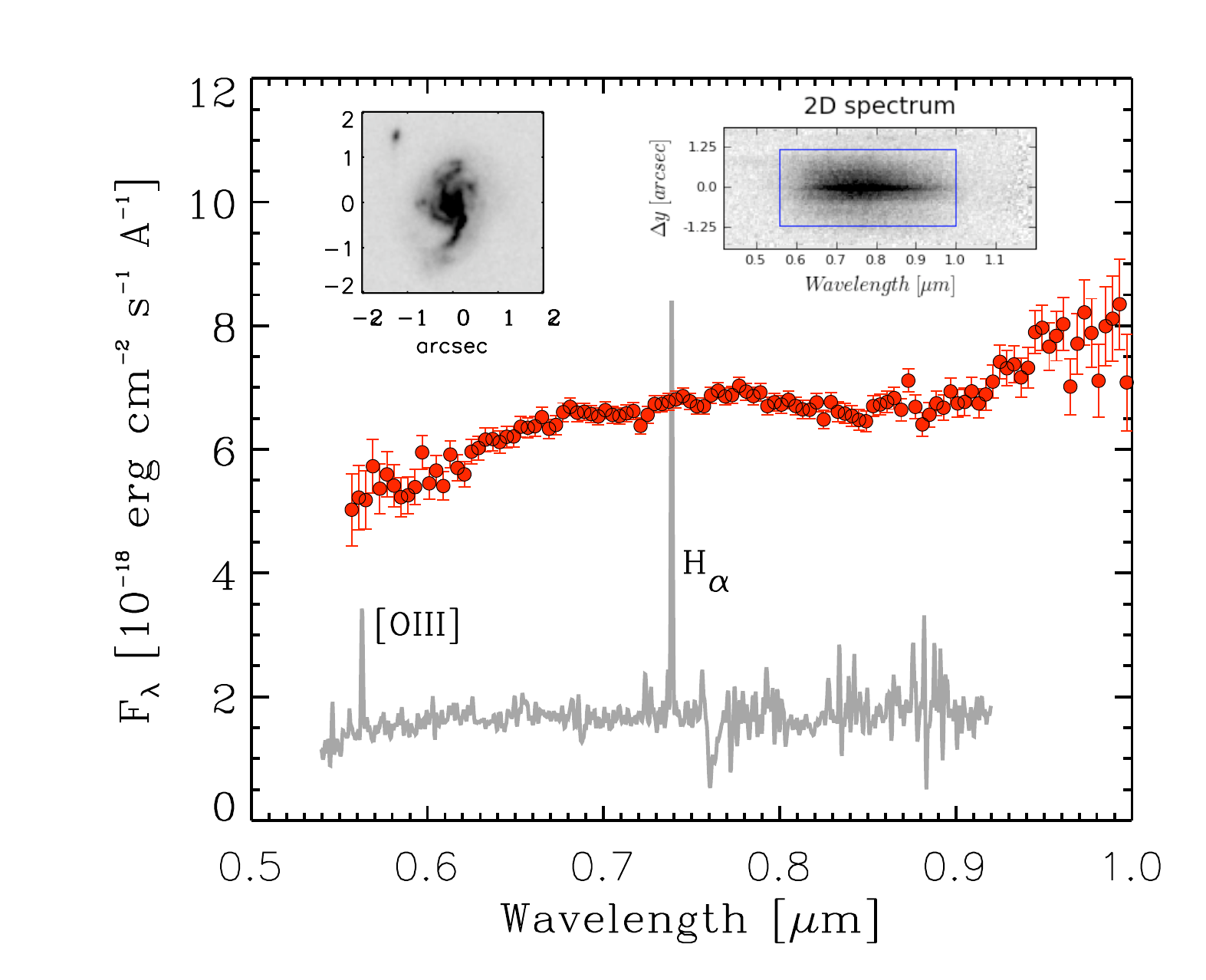}
\caption{Comparison of the ACS grism spectra (data points with error bars) 
with ground-based spectra (gray lines). Upper panel: the source
HAG\_J033209.44-274806.8 identified as a broad absorption line QSO
at z=2.81 by \cite{Szokoly04} using VLT/FORS2 spectroscopy with the 150I
grism (see Fig.\ \ref{fig:examples} for the full preview image).
Lower panel: The source HAG\_J033239.83-275652.8, a  spiral galaxy
at z=0.125, identified with a VLT/FORS2 spectrum taken with the 300I grism
as part of the ESO-GOODS survey \citep{Vanzellla08}. The insets show
the direct image stamps and, in the lower panel, also the 2D slitless stamp
image.}\label{fig:comp1}
\end{figure}

\begin{figure}
\includegraphics[width=\hsize]{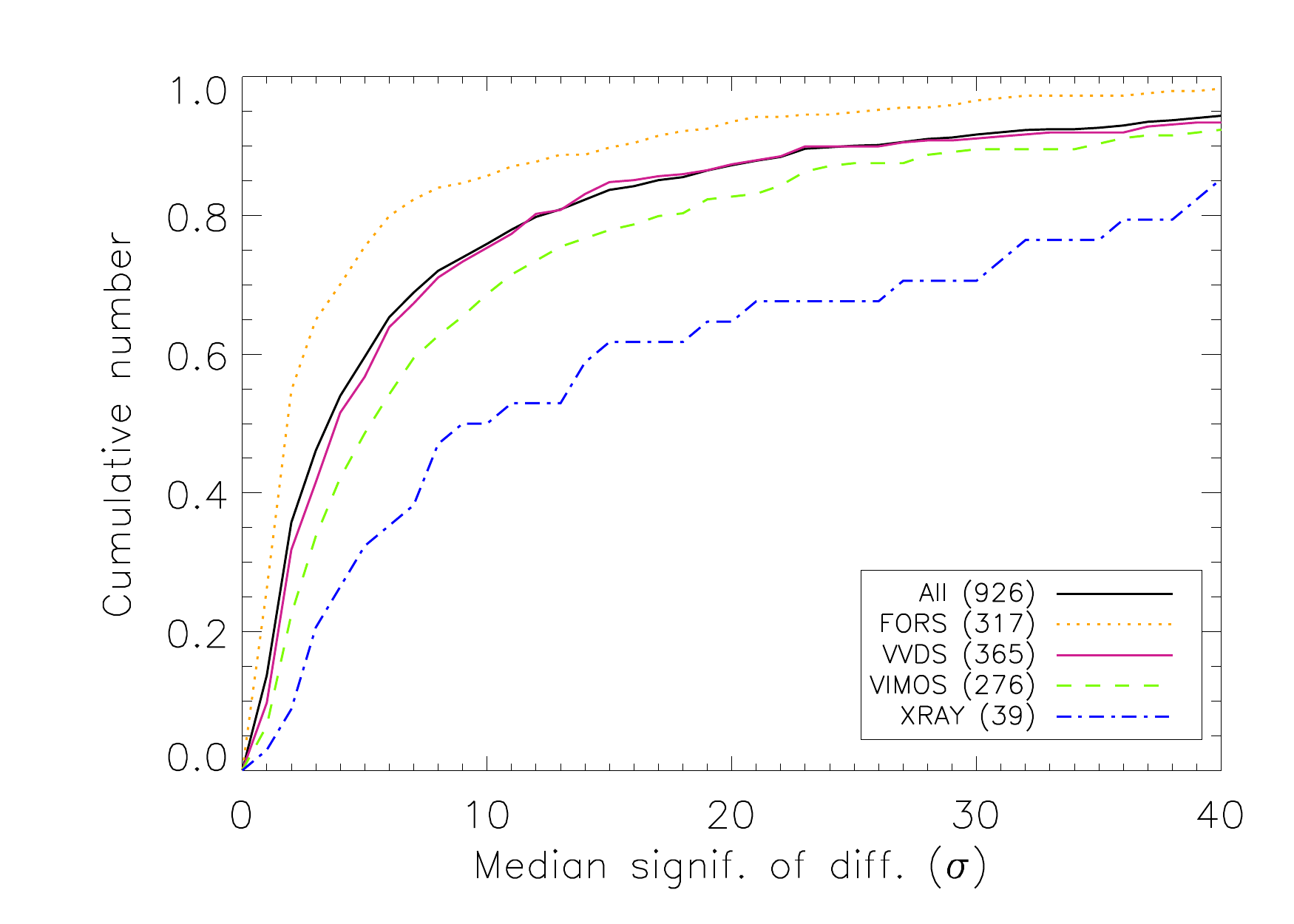}
\caption{Median $\sigma$ of all wavelength points of matched ground-based
and grism spectra, excluding points with large contamination. Black line shows
the distribution of all 926 spectra, and the coloured lines mark the various sub-samples
(see text). 92 spectra have a median significance $>25\sigma$. }
\label{fig:contcomp}
\end{figure}

For this test, the difference between all the FORS and VIMOS
spectra, rebinned and convolved with the resolution profile of the
grism spectrum, and their corresponding grism spectra were compared to
the combined errors on the grism spectra and on the ground-based
spectra, as calculated by the function \texttt{DER\_SNR}
\citep{Stoehr08}. All points in the spectra with contamination less
than 10\% were used. The median significance of this difference are
shown in Fig.~\ref{fig:contcomp} for the total sample of 926 matched
spectra. The figure also shows the same result divided into four
sub-samples of spectra; the GOODS-S FORS spectra, the VVDS VIMOS
spectra, the GOODS-S VIMOS spectra and the FORS spectra of CDFS X-ray
sources. Of the 926 matched spectra, 14\% have median significances
better than $1\sigma$, 46\% are better than $3\sigma$ and 76\% are
better than $10\sigma$. Considering that no corrections are made to
match or correct aperture sizes of the slitless and slit spectra, or
other systematic effects due to spectral feature, the agreement is
reasonably good.
Two examples are shown in Fig.\ \ref{fig:comp1}. The upper panel shows
the UV rest-frame of a quasar whose broad lines are clearly detected
in the grism spectrum. Also the overall flux levels agree reasonably well
with each other. In the case of the low-redshift spiral galaxy
(lower panel), the narrow emission lines (unresolved in the ground-based
spectrum with $R=660$) remain undetected in the grism spectrum, whereas
the mismatch in the continuum flux is the result of aperture effects.
As can be seen in the inset showing the 2D grism stamp plus the
extracted region (blue box), the slitless aperture is with
$2.5''$ much larger than the $1''$ used for the FORS2 slit spectrum.

\begin{figure}
\includegraphics[width=\hsize,clip=true]{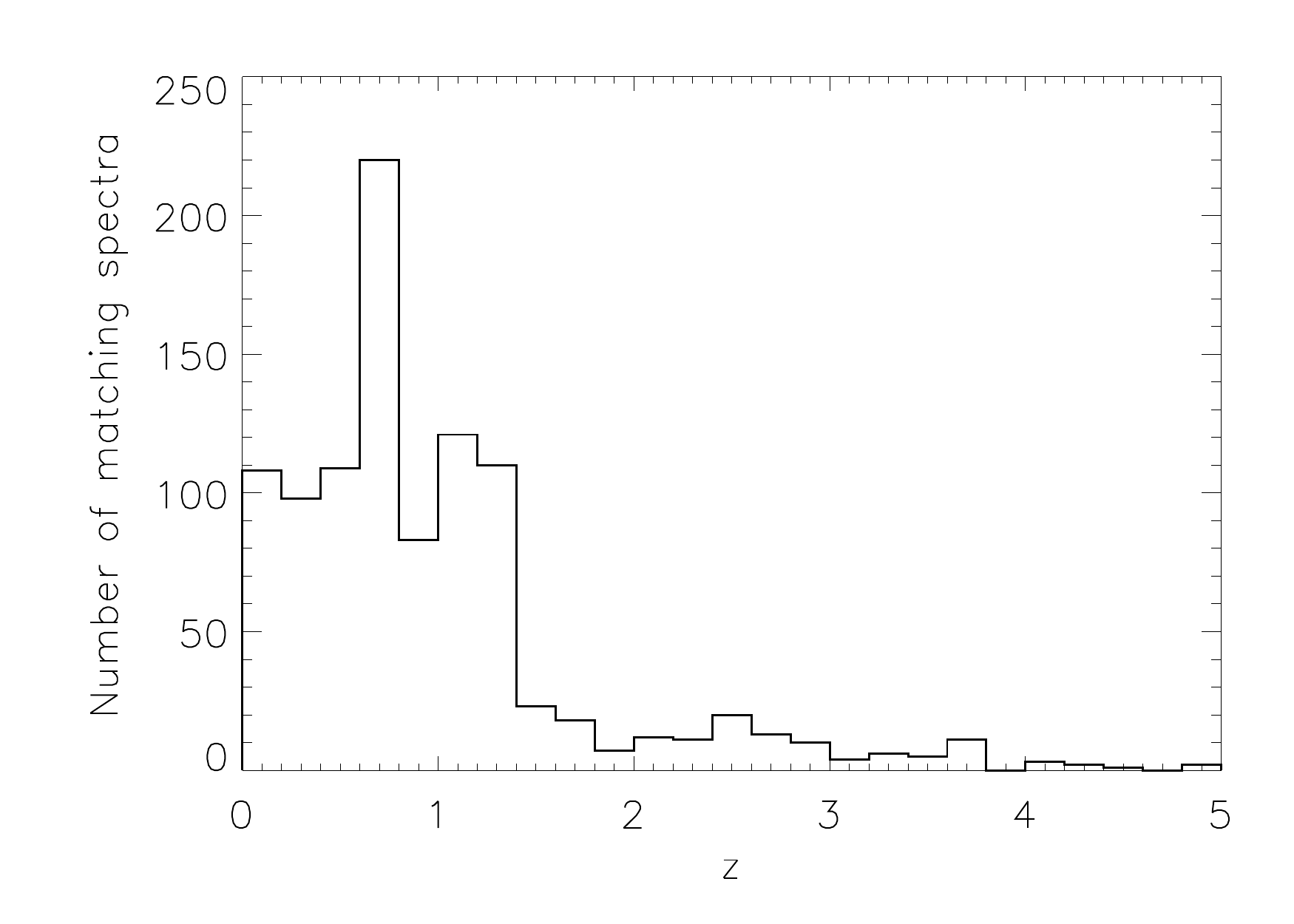}
\caption{Redshift distribution of grism sources with ground-based
  spectra from the FORS or VIMOS surveys.}
\label{fig:emlinez}
\end{figure}
\section{Empirical characterization of emission line sensitivities}
\label{sec:data_char}
%{[\bf Kim]}
We have seen how slitless spectroscopy is affected by contamination,
the spectral resolution is generally low and is further degraded for
extended sources, depending on their size, all aspects that differ
from typical slit spectroscopy and make a direct assessment of
limiting sensitivities difficult. An assessment of the flux continuum
and line limits, redshift accuracy etc. can be made by resorting to
well-studied samples of objects in common with ground-based spectra at faint
magnitudes in the GOODS South field.

Common targets were found in two steps.  Firstly,
matching objects were found within 1~arcsec radii. If multiple matches
were possible within this search radius, the closest match was
retained. Secondly, due to the overlap of several grism pointings, a
cleaning algorithm was applied to remove multiple entries of the same
source. This procedure produced $1,237$ matched sources distributed over
356 FORS spectra, 641 VIMOS spectra and 240 spectra from other
surveys. For this characterization, we will focus on the FORS and
VIMOS spectra. A plot of the redshift distribution of the FORS/VIMOS
matched sources can be found in Fig.~\ref{fig:emlinez}.
The distributions follow the overall distribution of ground-based
determined redshifts, i.e. the selection of objects in redshift
space was on average uniform and random. The drop in objects above
$z\sim 1.5$ is thus an artifact of declining galaxy number densities,
and spectroscopic incompleteness in the ground-based spectra.

\begin{figure}
\includegraphics[width=\hsize]{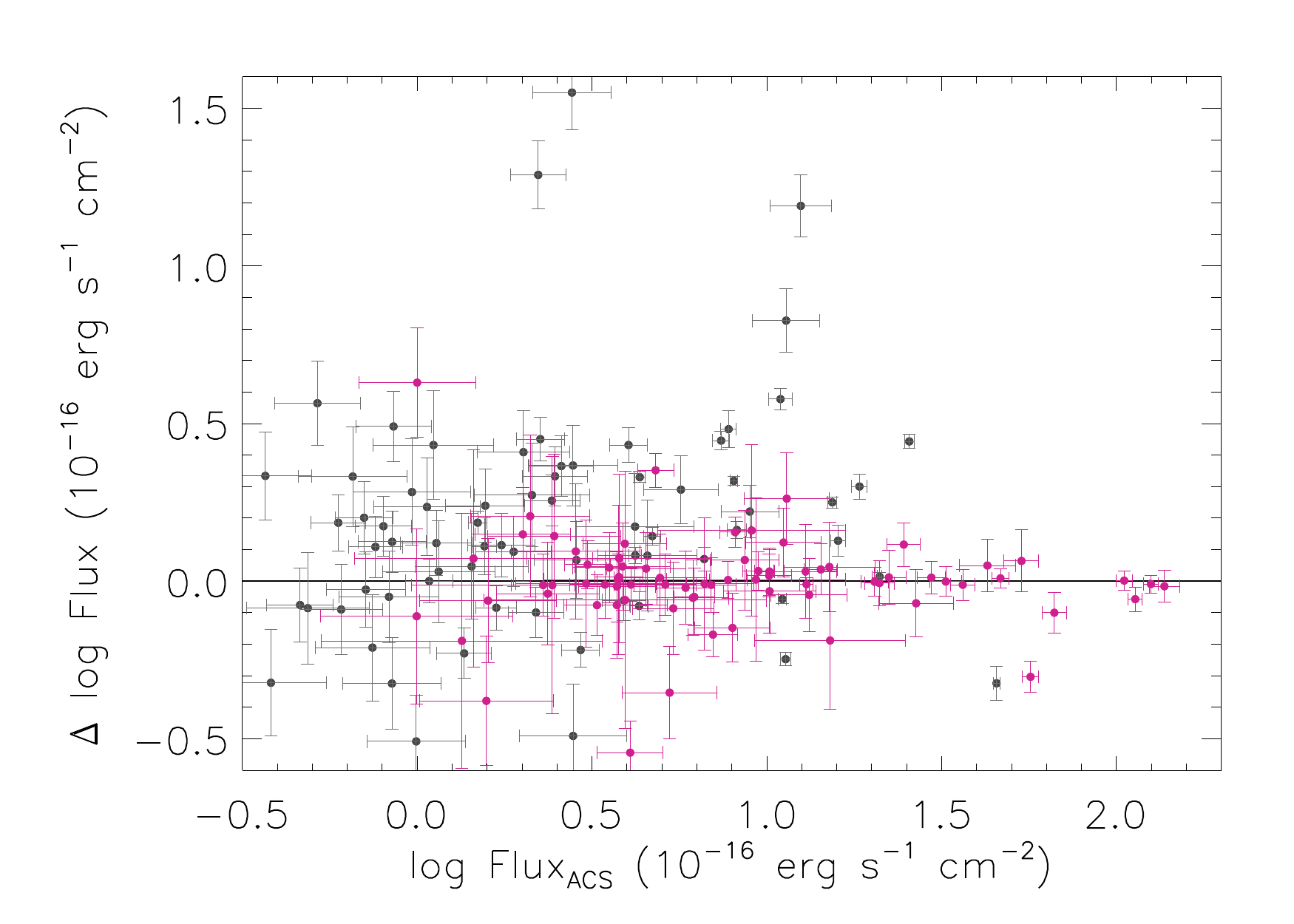}
\caption{Difference in the emission-line flux measured in the ground-based
(GB) and the grism spectra (ACS)
($\Delta \log{Flux} = \log{Flux_{GB}} - \log{Flux_{ACS}}$) as a function
of the ACS emission-line flux.
Grey/red points are before/after a tailored
  extraction. Only lines with 3$\sigma$ detections in both
  measurements are shown.  The solid line indicates a 1:1
  relation. Points above the solid line indicate too low flux
  measurements in the grism spectra. }
\label{fig:emlineresults}
\end{figure}

Before measuring emission-line fluxes, the
ground-based spectra were convolved with the PSF profile of the G800L
grism, also taking into account the extraction aperture size of the
corresponding grism spectrum. This ensures a matched resolution in the
ground-based and grism spectra. An automatic script was then used to
fit Gaussian functions to any emission-lines present in the wavelength
range $5500 - 9500$~{\AA}. The redshift was assumed to be that given
for the ground-based spectrum, and the line flux and EW were
calculated from the Gaussian fit. Measurements were only made in
regions were the contamination in the grism spectrum was less than
20\%.  Given the uncertainty in slit losses, the flux density in the
ground-based spectrum in the region around the line was normalized to
that in the grism spectrum. Due to these uncertainties, as well as the
effect of varying aperture sizes (see below), we focus here on the
emission-line fluxes, as opposed to equivalent width (EW).  The result
was stored and is shown as gray points in Fig.~\ref{fig:emlineresults}.

\begin{figure}[t]
\includegraphics[width=\hsize]{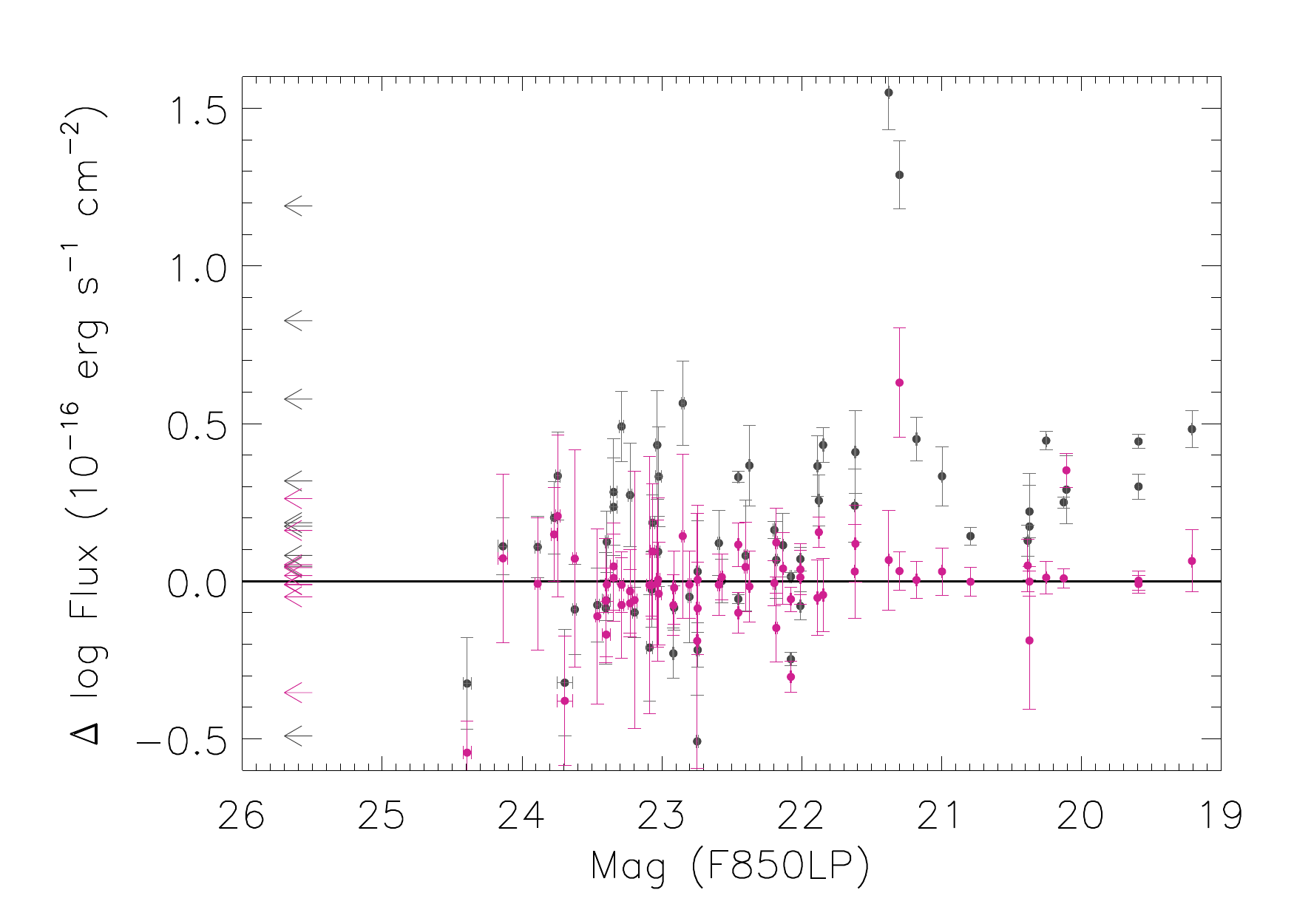}
\includegraphics[width=\hsize]{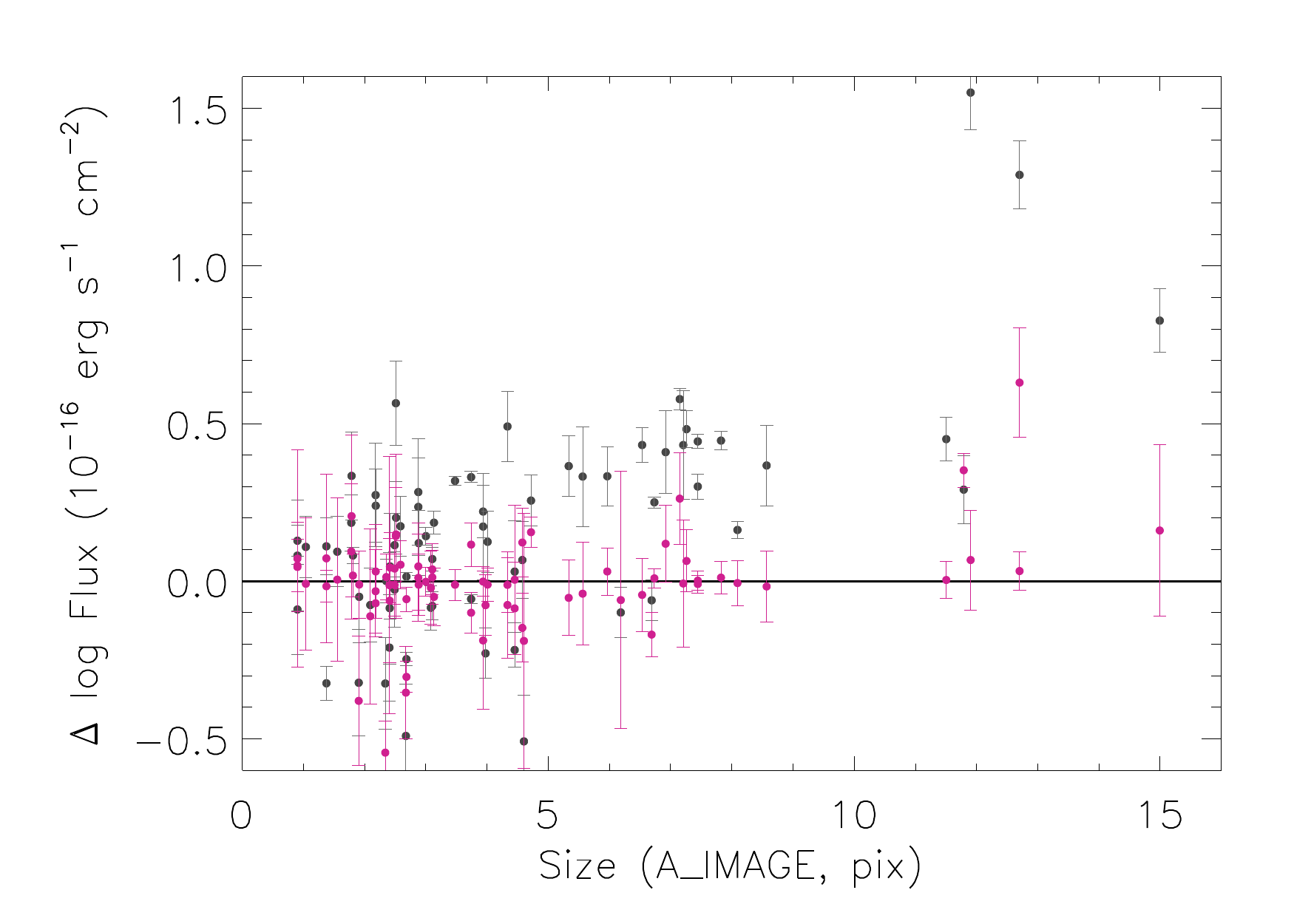}
\caption{Difference ($\Delta \log{Flux}$ as in Fig.\ \ref{fig:emlineresults})
in measured line fluxes as a function of overall magnitude in
the \emph{z} band (top) and major axis length as determined in the white
light image (bottom). Grey points are results before optimization and red
  after. Arrows in the upper plot indicate objects with upper limits on
  their magnitudes. The largest improvement is for large/bright
  objects, as expected. }
\label{fig:magsize}
\end{figure}

Even though a relation between the ground-based and grism fluxes is
seen, the agreement is not as good as expected. A reason for
a disagreement could be that emission-line regions can be small and
localized within larger galaxies. The pipeline will automatically set
an extraction aperture incorporating the whole galaxy, effectively
smearing the emission-line flux out. This would decrease the apparent
line flux compared to the ground-based measurement, as is apparent in
Fig.~\ref{fig:emlineresults}. To test this, and to compensate for it,
new one-dimensional spectra were extracted from the two-dimensional
extraction stamps in varying aperture sizes (see also
%\texttt{www.stecf.org/archive/hla/indiv\_extraction.php})
\texttt{http://hla.stsci.edu/STECF.org/archive/hla/} {\tt indiv\_extraction.php}.
The emission-lines in these spectra were then re-measured and the process
iterated until the best matches to the ground-based spectra were
found.  The results of this test are shown with red points in
Fig.~\ref{fig:emlineresults}, and display a much better correlation
between ground-based and grism measurements. Before the optimization,
58\% of the measured line fluxes disagreed with more than $2\sigma$,
the majority of those having too low fluxes as measured in the grism
spectra. After the optimization, the percentage of $>2\sigma$
discrepancies dropped to 9\%, with roughly equal number of objects
with too high/low flux as measured in the grism spectrum, consistent
with random scatter. It is clear that a refinement of the extraction is
necessary, particularly so in the case of more extended objects. In
Fig.~\ref{fig:magsize} the results are plotted as a function of size
and total magnitude of the objects.

Larger objects have, before optimization, the largest offsets between
ground-based and grism line measurements but are in good agreement
after optimization. Thus, the best improvement in the agreement is
achieved with larger objects. These objects also tend to be brighter
overall, and a strong improvement in the agreement is also seen for
bright objects in Fig.~\ref{fig:magsize}. It is important to note that
the line flux result after optimization was designed to match the line
flux of the ground-based spectrum, and is not necessarily the ``true''
line flux. The line flux/EW measured from a grism spectrum can vary
greatly, depending on extraction size, and location of extraction
centre. Thus, slitless spectra offer an opportunity to study spatially
resolved line emission for bright objects.

\begin{figure}
\includegraphics[width=\hsize]{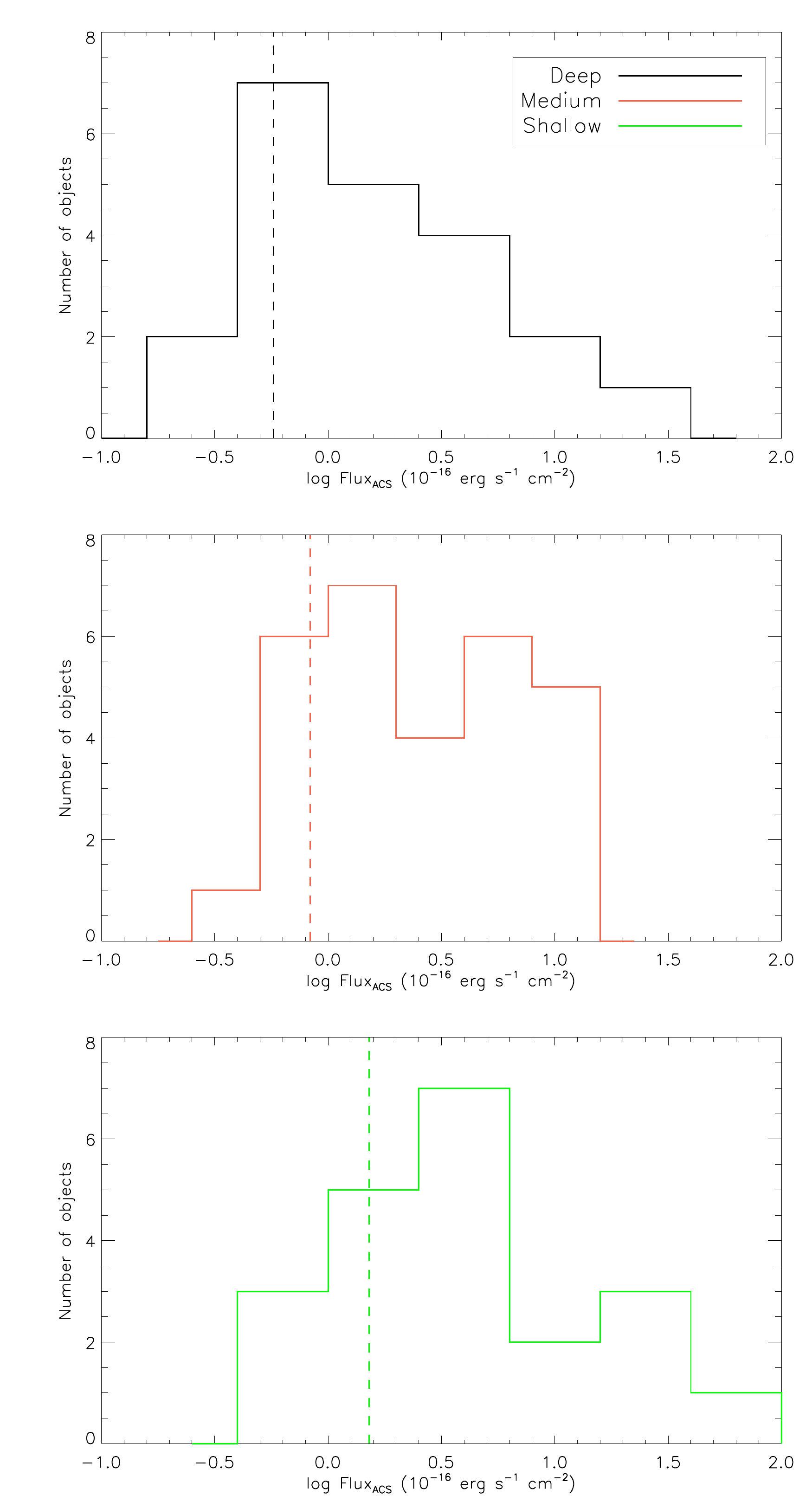}
\caption{Histogram of ACS line fluxes, from top to bottom as a function of
  decreasing exposure time. The median exposure times are $20,510$~s in the top
  panel, $15,030$~s in the middle panel and $4,320$~s in the bottom panel.
  The solid vertical lines mark the 80\%
  completeness limit in each sub-sample. }
\label{fig:exptime}
\end{figure}

The measurements made here can be used to calculate an empirical
sensitivity limit to emission lines in G800L observations. To study
this, the $3\sigma$ significant measurements were divided into three
bins depending on exposure time of the spectrum, with
$t_{exp}>16000$~s., $10000<t_{exp}\leq 16000$~s. and
$t_{exp}\leq10000$~s. The histogram of fluxes measured in the grism
spectra for these three bins are shown in Fig.~\ref{fig:exptime}.
To calculate an empirical detection limit, the 80\% completeness limit
in each sub-sample was found in the cumulative distribution. The
limits measured are $\log$~Flux$_{ACS} = (-16.24, -16.08,
-15.82)$~erg~s$^{-1}$~cm$^{-2}$ for the bins with mean exposure time
of $(20510, 15030, 4320)$~s. For these exposure times, and assuming a
Gaussian line with 100~{\AA} width and a S/N~$=3$, the limiting line
fluxes reached according to the ACS Exposure time calculator (ETC, see
http://etc.stsci.edu/etc/input/acs/spectroscopic/) are
$\log$~Flux~$ = (-16.70, -16.52,
-16.30)$~erg~s$^{-1}$~cm$^{-2}$. These values are significantly
fainter than the empirical limits found above, although the ETC values
depend on several factors, such as line EW, line central wavelength,
extraction size etc. If only the \emph{faintest} detected values in
Fig.~\ref{fig:exptime} are taken into account, the minimum fluxes
reached are $\log$~Flux$_{ACS} = (-16.43,
-16.31,-16.12)$~erg~s$^{-1}$~cm$^{-2}$, more in agreement with the ETC
values.

\begin{figure}
\includegraphics[width=\hsize]{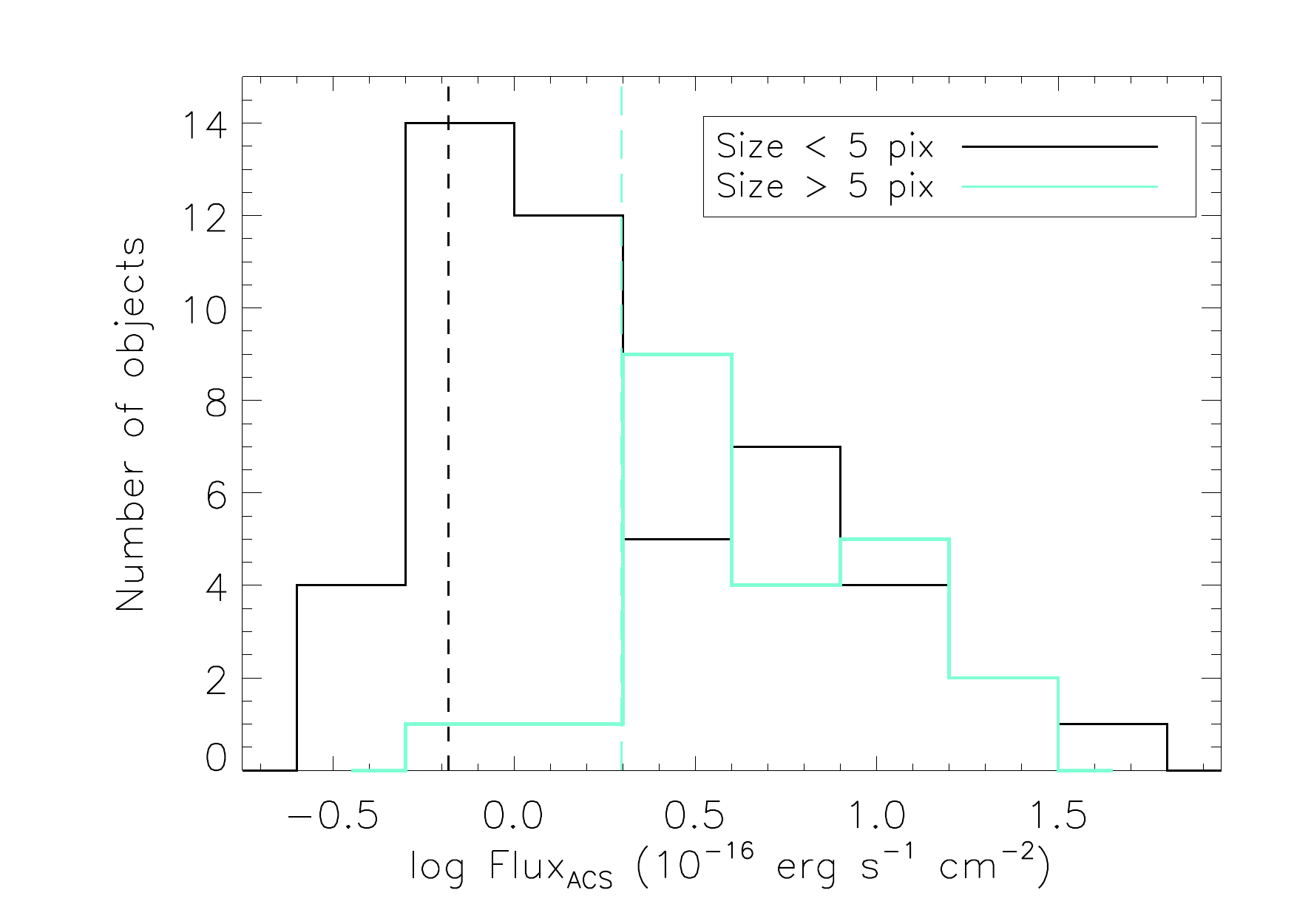}
\caption{Histogram of ACS line fluxes, divided into size bins
of {\tt A\_IMAGE} in pixel as measured on the detection image. Solid
  vertical lines mark the 80\% completeness limit in each
  sub-sample.}
\label{fig:sizematters}
\end{figure}

Finally, in
Fig.~\ref{fig:sizematters} the histogram of line fluxes is shown for
different size bins.
The results in the figure confirm that the limiting flux is size
dependent. The 80\% completeness in the two size sub-samples are
$\log$~Flux$_{ACS} = (-16.18, -15.70)$~erg~s$^{-1}$~cm$^{-2}$ for the
bins with sizes less/greater than five pixels.

\section{Data distribution}
HLA data are distributed by both ST-ECF and STScI and there are
several ways to access the ACS spectra:
\begin{itemize}
\item Archive Query Interface. HLA grism spectra can be searched on
  the dedicated HLA grism spectra interface
({\tt http://archive.eso.org/wdb/wdb/hla/} {\tt product\_science/form}) that
  allows many constraints on target, (e.g. the target name), the data
  properties (e.g. effective exposure time), the source properties
  (e.g. the magnitude) and the data quality (e.g. the signal-to-noise
  ratio). Also the general HST query interface at {\tt
    http://archive.eso.org/hst/science} includes all grism spectra,
  however with a less specific interface.

\item Archives at CADC and STScI. The ST-ECF HLA data has been also
  integrated into the general search interface at CADC
  ({\tt http://cadc.hia.nrc.gc.ca/hst/}) as well as into the HLA interface
  of the STScI ({\tt http://hla.stsci.edu}). The
  footprints of the equivalent slits of the grism spectra are also
  included into their footprint service.

\item Virtual Observatory. We provide fully automated access to the
  HLA metadata and data via Virtual Observatory (VO) standards. A
  Simple Spectrum Access Protocol (SSAP) server has been established
  ({\tt http://stecf.org/hst-vo/}).  It serves VOTables in V1.1
  format, which contain, in addition to the standard metadata,
  information about the footprints of the equivalent slits of the
  grism spectra. Our SSAP and SIAP server have been tested with ESO’s
  archive browser VirGO ({\tt http://archive.eso.org/cms/virgo/})
  as well as with SPLAT
  ({\tt http://star-www.dur.ac.uk/} {\tt $\sim$pdraper/splat/splat-vo/}) and VOSpec
    ({\tt http://esavo.esac.esa.int/vospec/}).
\end{itemize}

\section{General considerations for space-based slitless surveys}
Over the last decade, a number of space-based all-sky slitless surveys
have been designed with the aim of measuring baryonic acoustic
oscillations in the 3D-distribution of galaxies \citep[e.g.][]{Glaze05,Wang10}.
Space-based slitless experiments in the near-IR take
advantage of the low-background ($\sim1000$ times lower than on the
ground) and can cover wide areas with high sensitivity.
Galaxy redshift surveys conducted in this way can measure the large number
of redshifts ($>10^7$ emission line galaxies) needed to constrain the dark
energy equation of state parameter to a 1\% level, particularly in
the redshift interval $1\lesssim z\lesssim 2$, which is very hard to access
from the ground. Such experiments are part of the {\it Euclid} ESA
mission \citep{Lau09}, which is currently
in the definition phase, as well as of the {\sl JDEM} mission concept. 
%The experience accumulated with data from the slitless-modes of HST,
%such as the HLA data set described here, will be very useful when
%quantifying the performance of these future slitless experiments.

The large archive of ACS G800L spectra prepared in this work 
represents a unique collection of space-bourne slitless spectra and 
may prove useful in the planning of future slitless surveys from 
space. Expertise developed in 
the course of supporting ACS, and WFC3, slitless spectroscopy and 
the uniform reduction of ACS G800L spectra, described here, fed 
directly into simulations for the NIR slitless survey with 
Euclid, which featured in the Euclid Study Report ("The Yellow 
Book"; ESA 2009). 

ACS G800L data has a spectral resolution of about 80 for point 
sources and is adequate for redshift determination of emission 
line sources, spectral typing of stars and even modelling of 
late-type galaxy spectra \citep[e.g.][]{Ferreras2009}.
WFC3 NIR grisms have higher spectral resolution; the 2.6 times 
larger projected pixel size compared to ACS present some advantages 
in determining the redshift of single small ($\leq0.3 ''$) objects, 
but not, for example, for disentangling the spectra of larger 
objects with complex structure (such as late-type galaxies with 
HII regions). WFC3 NIR grism surveys published over the next few years will 
however be very useful for projecting the performance of future 
dark energy surveys that will be operating in the 1-2$\mu$m 
range and use primarily H$\alpha$ as a redshift indicator
\citep[e.g.][]{hakim}.

As emphasized in Section \ref{intro}, the contamination of spectra
is one of the 
most serious disadvantages of slitless spectroscopy as compared to 
multi-object slit spectroscopy. Its effect requires to be 
mitigated as much as possible in order to ensure a good return 
of clean spectra from a given slitless survey. The exact level 
of contamination depends on the target density, thus on the
Galactic latitude at the largest scale, and the length of the 
spectra on the detector. Other considerations that influence the 
contamination are the presence of multiple orders (including zeroth
order, where the similarity to an emission line for a point source
can be a significant hindrance in surveys relying on emission line
redshifts) and the relative strength of these 
orders with respect to the primary order (usually that carrying 
the most power). In the case of the ACS G800L, some contamination 
seriously affected around 40\% of the extracted spectra.
The ubiquity of contamination in slitless spectroscopy implies 
that strategies for dealing with contaminated spectra are a 
necessary part of the reduction process and should be folded 
into the analysis. 

For HST slitless surveys, the aim has been to provide as accurate 
as possible assessment of the likely effect of contamination on 
any given spectrum. The method developed for HST slitless spectra 
has relied upon morphological and spectral information from 
direct imaging. A direct image is mandatory for defining the
wavelength zero point for each extracted spectrum and plays a 
critical role in defining the size of the extraction (the 
virtual slit) and the effective resolution of a given slitless 
spectrum set by the object size projected in the dispersion 
direction. Thus one direct image is the minimum requirement 
for the slitless spectroscopy, but the availability of more 
than one image improves the contamination estimate by providing 
colour(s) of targets and a crude SED. Drop-out objects 
(containing a continuum break) are a particular case in point, 
since with only one image their contaminating flux must be, 
by default, assumed to be constant in flux and thus the extent of the 
contamination will be over-estimated. The use of the surface 
brightness distribution for extended sources \citep[flux cube 
contamination --][]{Kuemmel09b} rather than a 2D 
Gaussian fit, adds to the fidelity of the contamination 
estimation and directly to the yield of spectra with 
well-described contamination. The degree to which
contamination can be well described clearly plays an
important role in the fidelity of redshift determination.

The fraction of uncontaminated spectral pixels for a given
object clearly rises with the number of distinct rolls angles. 
For ACS the GRAPES \citep{Pirzkal2004} and PEARS surveys both 
had a minimum of 3 position angles. This observational strategy 
was found to allow the fraction of contaminated spectra 
(defined as below some level for a given fraction of pixels) to be 
approximately halved with four roll angles. The question of how to 
combine slitless spectra at different roll angle is however a complex 
one: for extended objects the spectra can differ radically depending 
on the projection of the virtual slit and cannot be simply averaged. 
Indeed the GRAPES and PEARS surveys distributed the spectra at 
different rolls angles as separate files. Only in the case of 
point or marginally extended sources should slitless spectra be
averaged. For all targets this is really only justified on the
basis of a given scientific analysis. Even for point sources, 
spectra from different rolls can contain different mixes of 
contaminating spectra, so combination after contamination removal 
must be contemplated before combination of spectra.

The flow diagram in Fig.\ \ref{fig:flchart} shows that 35\% of the
total number 
of extracted spectra were rejected in the course of the learning 
process. Conservative estimates suggest that this would fall to 
about half with four rolls angles, but general conclusions for other 
instruments must depend on dispersion, number and strength of orders, 
spatial pixel scale, etc. Detailed simulations using the available 
tools \citep[such as aXeSIM,][]{Kuemmel09b}, for typical fields 
are required to determine the exact return from multiple rolls, 
as was performed for the Euclid survey \citep[e.g.][]{geach}.

\section{Summary and Conclusion}
As part of the effort to enhance the content of the Hubble archive,
we have extracted slitless spectroscopic spectra taken with the ACS/WFC G800L
grism in $153$ archival fields which are distributed over the entire sky.
Using a fully automated data processing pipeline, we extracted $73,581$
spectra with a wavelength of $0.55$--$1.00 \mu\mbox{m}$ and a dispersion
of $40 \ \AA / \mbox{pixel}$ from the grism images. An automatic classification
algorithm, trained by the visual inspection of a test sample, was used for
the quality control. This algorithm selected $47,919$ ``good'' spectra for
publication that were derived from $32,149$ unique objects with a median 
$i_{\rm AB}$-band magnitude of $23.7$. Every released 1D spectrum is accompanied
with a 2D grism stamp spectrum, a direct image cutout in the available bands
and a preview image summarizing the results. All data is distributed
through a dedicated query interface and through VO access protocols.

The released grism spectra are science-ready, and many 
science cases can be addressed with these spectra. Slitless spectra
have been shown to be successful tools in finding and studying Galactic
stars \citep[e.g.][]{Pirzkal2009}. They also can be used to 
study emission-line galaxies such as described in Sect.~\ref{sec:data_char}
and Appendix \ref{sec:mult_line_em}
\citep[see also e.g.][]{Malhotra2005,Straughn,Straughn11}. 
Other higher redshift science cases involve
studying continuum features of high redshift galaxies such as the UV slope at
$z > 3.5$. \cite{nilsson} show the fitting of the spectral energy
distributions of a set of low redshift Lyman-break ($z\sim1$)
galaxies combining both photometric data points and ACS slitless spectra.

The experience accumulated with data from the slitless-modes of HST,
such as the HLA data set described here, will be very useful when
quantifying the performance of future space based slitless experiments.

\begin{acknowledgements}

  This paper is based on observations made with the NASA/ESA Hubble
  Space Telescope, obtained from the data archive at the Space
  Telescope -- European Coordinating Facility.  We thank our HLA
  collaborators Brad Whitmore and the STScI and CADC HLA teams.

\end{acknowledgements}

%\clearpage

%\clearpage

\appendix

\section{Multiple line emitters}
\label{sec:mult_line_em}
If no other information is available for a celestial object, typically
two or more emission-lines observed in a spectrum will suffice to
determine the redshift of the source. With this goal in mind,
i.e.~determining redshifts for objects, all grism spectra were
searched for double or multiple emission lines within the same
spectrum. Note that it is not in the scope of this publication to create an 
exhaustive list of emission-line redshifts. This section is merely meant as
an introduction to the method, and interested users may find more objects
using other selection criteria. The criteria for selection for follow-up
used here were the following:
\begin{itemize}
\item[1:] Only the wavelength region $5600$--$9500$~{\AA} was
  considered.
\item[2:] Only spectral points with contamination less than 10\% of the flux were considered.
\item[3:] Only spectral points with flux greater than $4\sigma$ significance 
\emph{after subtraction of the continuum} (fitted with a second degree polynomial) were considered.
\item[4:] At least three adjacent spectral points satisfying criterion $3$ had to be found in 
each of the two wavelength regimes $5700 - 7400$~{\AA} and $7600 - 9300$~{\AA}.
\item[5:] A visual inspection adjusted the parameters above to the optimal selection, i.e. minimising false detections,
and rejected some bad spectra.
\item[6:] Finally, spectra of duplicate objects (in different associations) were excluded 
by matching the locations within $1''$.
\end{itemize}
A selection with these criteria will allow detection of strong line emitters in the
redshift ranges of $z = 0.158 - 0.417$ for the combination H$\beta$/[OIII]/H$\alpha$,
$z = 0.529 - 0.857$ for [OII]/H$\beta$/[OIII] and $z \sim 3.9 - 5.0$ for Ly$\alpha$/CIV.
The selected sample consisted of 54 objects. For these 54 the obvious 
emission-lines were fitted with Gaussian functions to retrieve the central 
wavelength of the lines. From these central wavelengths, redshifts were 
determined. We report these determinations, including spectral IDs and line 
identifications in Table~\ref{tab:redshifts}.

The redshift and its significance intervals reported in the Table are
determined from the central wavelengths (and uncertainties thereof) of
the H$\alpha$ or the [OII] lines when available. For the Ly$\alpha$
emitters the redshift is the mean from all measured lines in the
spectrum and the uncertainty is the standard deviation on those
measurements. Note that the uncertainties are in the fitting of
the Gaussian line, and do not include systematic uncertainties due to
e.g. the sizes of the galaxies. Thus they must be considered lower
estimates of the uncertainties.  Allowing a potential systematic error
of 1 spectral pixel (40~{\AA}) would for instance result in a $\delta
z =$~[0.006, 0.011, 0.033] for [H$\alpha$, [OII], Ly$\alpha$].  It may
also be possible for users to find more multiple line emitters, if
allowing smaller significances on the emission line fluxes (see
criterion 3 above), larger spectral intervals (criteria 1 and 4
above) or relaxing the contamination criterion. 

9 of the objects in Tab.\ \ref{tab:redshifts} are located in the
GOODS fields, and around $\sim80\%$ of them have already published
redshifts. Outside of the GOODS fields only four sources have already
published redshifts. The four sources with large differences between
our redshift estimates and the published values are discussed in the
Notes to Tab.\ \ref{tab:redshifts}.

\begin{table*}[t]
\centering
\caption{List of failed associations}
\begin{tabular}{@{}ccccc}
\hline
\hline
Association& RA (2000)& DEC (2000)& $n_{grism}$& Note\\
\hline
J8HP9SBMQ& 23:02:33.54& 8:56:26.6& 3&  (1)\\
J8HP9TGBQ& 0:43:47.68& 40:46:51.4& 5& (2)\\
J8HP9VT2Q& 0:43:40.77& 40:44:03.0& 5&  (2)\\
J8HP9XKZQ& 0:43:38.76& 40:41:59.9& 1&  (3)\\
J8HPAJQGQ& 11:29:35.78& -14:41:15.8& 1&  (3)\\
J8HPDMLQQ& 15:42:41.23& -11:02:33.7& 5& (1)\\
J8HPDNVTQ& 11:18:55.56& 7:40:10.6& 2&(1)\\
J8HQ9VMQQ& 13:24:49.16& 57:06:18.7& 3& (1)\\
J8HQDPGZQ& 10:03:38.98& 29:05:27.7& 1& (3)\\
J8HQE9P2Q& 9:55:15.77& 17:33:20.0& 5& (1)\\
J9R73IAWQ& 11:56:27.89& 54:57:01.3& 16& (4)\\
J9R73MLDQ& 11:56:15.16& 54:57:40.3& 16& (4)\\
J9R76MMKQ& 9:50:41.22& 33:24:16.1& 16& (4)\\
J9R7A2EBQ& 3:22:08.12& -15:34:24.8& 14&	 (4)\\
J9R7B1AUQ& 0:46:40.90& 17:08:39.9& 8& (4)\\
\hline
\end{tabular}
\begin{minipage}{\textwidth}
\begin{tabular}{lll}
Notes: & (1) & Insufficient direct images\\
       & (2) & Object density too high\\
       & (3) & Not sufficient grism images\\
       & (4) & Bad background subtraction for grism images\\
\label{tab:failedassocs}
\end{tabular}
\end{minipage}
%\end{center}
\end{table*}

\begin{table*}[t]
\label{tab:redshifts}
\begin{center}
\caption{Objects with determined redshifts}
\begin{tabular}{@{}lcccccccccc}
\hline
\hline
HLA Object ID & RA & Dec & $z$ & Line IDs \\
\hline
HAG\_J022719.28-405741.2\_J8HQBUUYQ & 36.83033 & -40.96145 & $0.349\pm0.002$ & H$\beta$/[OIII]/H$\alpha$ \\
HAG\_J033237.80-275606.1\_J9FA43BIQ & 53.15750 & -27.93502 & $0.331\pm0.001$ & H$\beta$/[OIII]/H$\alpha$ \\
HAG\_J033238.04-275508.1\_J9FA4FXBQ & 53.15850 & -27.91891 & $0.370\pm0.001$ & H$\beta$/[OIII]/H$\alpha$ \\
HAG\_J033236.16-275408.8\_J9FA43BIQ & 53.15066 & -27.90244 & $0.277\pm0.001$ & H$\beta$/[OIII]/H$\alpha$ \\
HAG\_J033239.72-275154.6\_J9FA63NNQ & 53.16551 & -27.86517 & $0.393\pm0.016$ & H$\beta$/[OIII]/H$\alpha$ \\
HAG\_J033244.27-275141.1\_J94SP9ACQ & 53.18448 & -27.86141 & $0.277\pm0.001$ & H$\beta$/[OIII]/H$\alpha$ \\
HAG\_J033217.56-274941.0\_J94SA3CIQ & 53.07318 & -27.82805 & $0.341\pm0.009$ & H$\beta$/[OIII]/H$\alpha$ \\
HAG\_J033237.11-274735.6\_J8QQ34PZQ & 53.15464 & -27.79323 & $0.230\pm0.002$ & H$\beta$/[OIII]/H$\alpha$ \\
HAG\_J033246.97-274715.8\_J8QQ10IKQ & 53.19572 & -27.78774 & $0.227\pm0.001$ & H$\beta$/[OIII]/H$\alpha$ \\
HAG\_J033234.74-274707.6\_J8G6I3PEQ & 53.14474 & -27.78544 & $0.239\pm0.001$ & H$\beta$/[OIII]/H$\alpha$ \\
HAG\_J033222.67-274403.0\_J9FA83N9Q & 53.09444 & -27.73417 & $0.313\pm0.001$ & H$\beta$/[OIII]/H$\alpha$ \\
HAG\_J033225.91-274401.5\_J9FA83N9Q & 53.10798 & -27.73376 & $0.279\pm0.001$ & H$\beta$/[OIII]/H$\alpha$ \\
HAG\_J033215.83-274351.3\_J9FAA3IBQ & 53.06596 & -27.73092 & $0.362\pm0.002$ & H$\beta$/[OIII]/H$\alpha$ \\
HAG\_J033219.80-274122.8\_J9FAA3IBQ & 53.08252 & -27.68967 & $0.220\pm0.001$ & H$\beta$/[OIII]/H$\alpha$ \\
HAG\_J010616.13-274101.4\_J8HQ9ETYQ & 16.56720 & -27.68372 & $2.211\pm0.004$ & [CIII]/MgII \\
HAG\_J013014.65-160257.5\_J8HPCAYEQ & 22.56106 & -16.04930 & $0.280\pm0.001$ & H$\beta$/[OIII]/H$\alpha$ \\ 
HAG\_J053003.17-071438.5\_J8HPAKG9Q & 82.51321 & -7.24403 & $4.19\pm0.08$ & Ly$\alpha$/SiII/SiIV/CIV \\
HAG\_J021844.48-044824.7\_J6FL8XDSQ & 34.68533 & -4.80685 & $4.55\pm0.01$ & Ly$\alpha$/SiIV/CIV \\
HAG\_J021849.90-044725.9\_J6FL7XOYQ & 34.70793 & -4.79052 & $0.630\pm0.003$ & [OII]/H$\beta$/[OIII] \\
HAG\_J134002.60+000916.9\_J8HPB3GFQ & 205.01083 & 0.15469 & $3.88\pm0.02$ & Ly$\alpha$/SiIV/CIV \\
HAG\_J104726.77+053416.3\_J8SB10CEQ & 161.86154 & 5.57119 & $0.248\pm0.001$ & H$\beta$/[OIII]/H$\alpha$ \\
HAG\_J091142.77+055751.7\_J8HPBNA4Q & 137.92819 & 5.96435 & $0.19\pm0.01$ & H$\beta$/[OIII]/H$\alpha$ \\
HAG\_J080859.40+064227.3\_J8IY3LHOQ & 122.24750 & 6.70759 & $0.322\pm0.002$ & H$\beta$/[OIII]/H$\alpha$ \\
HAG\_J080857.79+064233.8\_J8IY3LHOQ & 122.24077 & 6.70939 & ? & Lines at $8158, \, 8599$~{\AA} \\
HAG\_J220358.19+184907.4\_J8HQDKAFQ & 330.99247 & 18.81872 & $3.987\pm0.007$ & Ly$\alpha$/SiIV/CIV/HeII \\
HAG\_J144555.47+403625.4\_J8HP9DU4Q & 221.48114 & 40.60705 & $4.12\pm0.07$ & Ly$\alpha$/SiII/SiIV/CIV \\
HAG\_J004527.48+403922.6\_J8HPDCAOQ & 11.36450 & 40.65628 & $0.295\pm0.001$ & H$\beta$/[OIII]/H$\alpha$ \\
HAG\_J083248.98+523951.1\_J8HP9NN6Q & 128.20410 & 52.66420 & $0.247\pm0.001$ & H$\beta$/[OIII]/H$\alpha$ \\
HAG\_J100237.23+545554.3\_J8HQCBULQ & 150.65512 & 54.93175 & $0.770\pm0.006$ & [OII]/H$\beta$/[OIII] \\
HAG\_J123655.70+620823.5\_J8N1ZABEQ & 189.23210 & 62.13987 & $0.219\pm0.001$ & H$\beta$/[OIII]/H$\alpha$ \\
HAG\_J123651.12+620938.7\_J8N1ZABEQ & 189.21298 & 62.16075 & $0.204\pm0.001$ & H$\beta$/[OIII]/H$\alpha$ \\
HAG\_J123704.29+620959.8\_J8N1ZABEQ & 189.26787 & 62.16660 & $0.318\pm0.001$ & H$\beta$/[OIII]/H$\alpha$ \\
HAG\_J123731.48+621005.9\_J94SX2FTQ & 189.38115 & 62.16830 & $0.169\pm0.001$ & H$\beta$/[OIII]/H$\alpha$ \\
HAG\_J123717.42+621046.8\_J8G6Z3JDQ & 189.32259 & 62.17967 & $0.61\pm0.01$ & [OII]/H$\beta$/[OIII] \\
HAG\_J123720.38+621047.2\_J94SB2A1Q & 189.33492 & 62.17978 & $0.202\pm0.001$ & H$\beta$/[OIII]/H$\alpha$ \\
HAG\_J123636.87+621134.9\_J9FAGEPGQ & 189.15363 & 62.19304 & $0.078^1$ & Multiple spectral ``bumps'' \\
HAG\_J123650.82+621255.9\_J94SE2QUQ & 189.21174 & 62.21553 & $0.325\pm0.001$ & H$\beta$/[OIII]/H$\alpha$ \\
HAG\_J123658.06+621300.4\_J94SE2QUQ & 189.24193 & 62.21679 & $0.314\pm0.001$ & H$\beta$/[OIII]/H$\alpha$ \\
HAG\_J123810.10+621620.5\_J8WQI1BKQ & 189.54210 & 62.27235 & $0.313\pm0.003$ & H$\beta$/[OIII]/H$\alpha$ \\
HAG\_J123757.31+621627.4\_J8WQI1BKQ & 189.48879 & 62.27429 & $3.972^2\pm0.008$ & Ly$\alpha$/SiII/CIV/HeII \\
HAG\_J123752.72+621628.2\_J8WQI1BKQ & 189.46968 & 62.27449 & $0.306^3$ & --- \\
HAG\_J123752.74+621752.0\_J8WQI1BKQ & 189.46974 & 62.29777 & $0.302\pm0.002$ & H$\beta$/[OIII]/H$\alpha$ \\
HAG\_J123742.54+621811.8\_J9FAJLDCQ & 189.42724 & 62.30328 & $2.278\pm0.004$ & H$\beta$/[OIII]/H$\alpha$ \\
HAG\_J123717.89+621855.7\_J9FAJEBAQ & 189.32453 & 62.31547 & $2.213\pm0.004$ & [CIII]/MgII \\
HAG\_J123721.25+621915.4\_J9FAILOXQ & 189.33855 & 62.32094 & $0.236\pm0.001$ & H$\beta$/[OIII]/H$\alpha$ \\
HAG\_J123616.05+621927.8\_J8HQAAKAQ & 189.06690 & 62.32439 & $0.223\pm0.001$ & H$\beta$/[OIII]/H$\alpha$ \\
HAG\_J123612.11+621941.1\_J8HQAAKAQ & 189.05046 & 62.32809 & $2.214\pm0.003$ & [CIII]/MgII \\
HAG\_J123719.71+621943.2\_J8G6F1BTQ & 189.33211 & 62.32865 & $0.644^4\pm0.002$ & [OII]/H$\beta$/[OIII] \\
HAG\_J123725.07+622005.0\_J9FAJEBAQ & 189.35445 & 62.33471 & $0.276\pm0.001$ & H$\beta$/[OIII]/H$\alpha$ \\
HAG\_J123721.13+622019.6\_J8G6F1BTQ & 189.33804 & 62.33877 & $0.261\pm0.001$ & H$\beta$/[OIII]/H$\alpha$/[SII] \\
HAG\_J123713.47+622137.2\_J8WQF2BRQ & 189.30611 & 62.36034 & $0.291\pm0.001$ & H$\beta$/[OIII]/H$\alpha$ \\
HAG\_J135835.38+623858.1\_J8HQ9SKYQ & 209.64742 & 62.64946 & $0.703\pm0.001$ & [OII]/H$\beta$/[OIII] \\
HAG\_J163614.42+660543.6\_J8HQEFQRQ & 249.06007 & 66.09546 & $0.342\pm0.003$ & H$\beta$/[OIII]/H$\alpha$ \\
HAG\_J163604.14+661302.2\_J8ZE02NJQ & 249.01726 & 66.21727 & $0.284\pm0.001$ & H$\beta$/[OIII]/H$\alpha$ \\
\hline
\end{tabular}
\end{center}
\begin{list}{}{}
\item[\textbf{Notes.}]$^1$Redshift undetermined from our spectrum, redshift in table from Wirth et al.~(2004). $^2$ Barger et al.~(2002) reports a 
redshift of $z = 2.922$ for this source, but the our slitless spectrum appears to be of better quality and we retain our redshift determination. $^3$ Redshift from our grism
spectrum is $z_{gr} = 0.718\pm0.003$, but the redshift given in the table comes from a better quality spectrum of Wirth et al.~(2004). $^4$ For this source, Barger et 
al.~(2008) reports a redshift of $z = 0.247$. As this spectrum is not available for inspection, and the our spectrum is of good quality, we retain 
the redshift determined from our grism spectrum.
\end{list}
\end{table*}

\longtabL{3}{
\label{tab:assoc}
\begin{landscape}
\begin{longtable}{lrrrrrrlrrrr}
\caption{The associations included in this release. The column
{\tt Orient.}\ contains the orientation of the y-axis in the detection image.
The column
{\tt Flag} marks the image quality of the detection image (0: very good,
1: acceptable, 2: not optimal).}\\
\hline \hline
Association& RA (2000) & DEC (2000)& Gal. Lat. & Gal. Long. & Orient. & Obs-date & Filter,\,$n_{img}$,\,$expt_{img}$& $n_{grism}$ & $expt_{grism}$ & $N_{obj}$ & Flag\\ \hline
\endfirsthead
  J6FL7XOYQ&  2:18:51.01& -4:47:31.6&169.8086454& -59.4561658&         38.4&         2002-11-19& F850LP, 6, 3000.0& 8& 6112.0& 265& 1\\
  J6FL7YIGQ&  2:17:34.01& -4:53:51.4&169.4641039& -59.7441393&         28.0&         2002-11-25& F850LP, 6, 3000.0& 8& 6112.0& 273& 1\\
  J6FL8XDSQ&  2:18:51.01& -4:47:32.2&169.8088323& -59.4562769&         37.1&         2002-11-18& F850LP, 6, 3000.0& 8& 6112.0& 262& 1\\
  J6FL8YKUQ&  2:17:33.92& -4:53:51.1&169.4634487& -59.7443249&         33.0&         2002-11-26& F850LP, 6, 3000.0& 8& 6112.0& 256& 1\\
  J8G6F1BTQ& 12:37:26.16& 62:19:08.0& 125.747724&  54.7327956&       -133.0&         2003-03-03& F435W, 9, 10800.0& 12& 13900.0& 336& 0\\
                                                                                               & & & & & & &F606W, 22, 9904.0& & & & \\
                                                                                               & & & & & & &F775W, 23, 10410.0& & & & \\
                                                                                               & & & & & & &F850LP,17, 10100.0& & & & \\
\end{longtable}
\end{landscape}
}

\end{document}